\documentclass[aps,pre,showpacs,superscriptaddress]{revtex4}
\usepackage{graphicx}

\newcommand{\la}{\lambda}
\newcommand{\ep}{\epsilon}

\newcommand{\xv}{\vec{x}}
\newcommand{\be}{\begin{equation}}
\newcommand{\ee}{\end{equation}}
\newcommand{\bea}{\begin{eqnarray}}
\newcommand{\eea}{\end{eqnarray}}
\newcommand{\ad}{a^{\dagger}}
\newcommand{\V}[1]{{\bf #1}}
\newcommand{\pd}[2]{\frac{\partial #1}{\partial #2}}
\newcommand{\dd}[2]{\frac{d #1}{d #2}}

\begin{document}

\title{Cluster-Cluster Aggregation as an Analogue of a Turbulent Cascade :
Kolmogorov Phenomenology, Scaling Laws and the Breakdown of Self-Similarity}
\author{Colm Connaughton}
\email{colm@cnls.lanl.gov}
 \affiliation {Laboratoire de Physique Statistique de l'ENS, 
24 rue Lhomond - 75231 Paris cedex 05, France}
 \affiliation {Current address : Centre for Nonlinear Studies, Los Alamos 
National Laboratory, Los Alamos, NM 87545, U.S.A.}
 \author{R. Rajesh}
\email{rrajesh@imsc.res.in}
\affiliation{The Institute of Mathematical Sciences, CIT Campus, Taramani, 
Chennai 600 113, India}
 \author{Oleg Zaboronski}
\email{olegz@maths.warwick.ac.uk}
 \affiliation{Mathematics Institute, University of Warwick, Gibbet Hill
Road, Coventry CV4 7AL, UK}

\date{\today}

\begin{abstract}
We present a detailed study of the statistical properties of a system of 
diffusing aggregating particles in the presence of a steady source of monomers.
We emphasise the case of low spatial dimensions where strong diffusive 
fluctuations invalidate the mean-field description provided by standard
Smoluchowski kinetic theory. The presence of a source of monomers allows
the system to reach a statistically stationary state at large times. This
state is characterised by a constant flux of mass directed from small to large
masses. It therefore admits a phenomenological description based on the 
assumption of self-similarity and constant mass flux analogous to the 
Kolmogorov's 1941 theory of turbulence. Unlike turbulence, the aggregation
problem is analytically tractable using powerful methods of statistical
field theory. We explain in detail how these methods should be adapted to
study the far-from-equilibrium, flux-dominated states characteristic of 
turbulent systems. We consider multipoint correlation functions of the
mass density. By an exact evaluation of the scaling exponents for the one and 
two-point correlation functions, we show that the assumption of 
self-similiarity breaks down at large masses for spatial dimensions, $d\leq2$. 
We calculate non-rigourously the exponents of the higher order correlation
functions as an $\epsilon$-expansion where $\epsilon=2-d$. We show that the 
mass distribution exhibits non-trivial multiscaling. An analogy can be drawn
with the case of hydrodynamic turbulence. The physical origin of this
multiscaling is traced to the presence of strong correlations between
particles participating in large mass aggregation events. These correlations 
stem from the recurrence of diffusion processes in $d\leq2$.  
The analytic methods developed here will have more general applicability beyond
the study of this specific problem. 
\end{abstract}


\maketitle

\section{Introduction}
\label{sec-introduction}
The development of a coherent formalism that enables us to calculate the 
statistical properties of complex systems far from equilibrium remains a vexing 
problem in theoretical physics. It may be that the set of non-equilibrium
complex systems is too diverse to admit a unified description analogous to
the Gibbs formulation for equilibrium systems. Nevertheless certain 
subsets of non-equilibrium systems share enough common features that it
is reasonable to hope that a general understanding can be obtained. One such
subset which has attracted much interest, both theoretical and practical,
over the years is the class of problems of turbulent type. In the
present context, turbulence does not refer exclusively to hydrodynamic
problems although the statistics of high Reynolds number fluid flow is
a very important and challenging example. Rather, we take the word turbulent
to refer to a class of non-equilibrium problems coming from a diverse range 
of areas in theoretical physics (from hydrodynamics  to condensed matter 
physics to aggregation and even cosmology) sharing certain common features 
which we now describe.

The defining characteristic of a turbulent system is the existence of a 
stationary state with widely separated sources and sinks 
of some conserved quantity. While stationary states of turbulent systems lack 
detailed balance, they are 
characterised by a flux, mediated by nonlinear interactions,
of a conserved quantity from source to sink. 
Universal statistics are expected  in
regions far from sources and sinks, also known as the inertial range. 
The best known 
example, as mentioned already, is the solution of the
Navier--Stokes (N--S) equations at high Reynolds number with large scale
forcing \cite{frischBook}.  It is characterised by an
energy flux from large length scales to small length scales
where viscosity dissipates the energy.
Other examples include Burgers turbulence \cite{falkovich2001}, the Kraichnan 
model of passive scalar advection \cite{falkovich2001}, 
kinematic magneto hydrodynamics \cite{falkovich2001},
and wave turbulence \cite{zakharovBook}.
A common feature of these systems is that they each 
admit a phenomenological description based on constancy of the flux and the 
assumption of self-similarity. The first such theory was the Kolmogorov 1941 
(K41) theory of N--S turbulence. Obtaining a quantitative understanding of 
the limitations  and applicability of such phenomenology is one of the major
theoretical challenges. Analytic progress in hydrodynamics has been slow 
despite strong 
numerical and experimental evidence for many interesting and nontrivial 
violations of K41-style self-similarity. This is because the N--S equations
lack an obvious small parameter that would permit a perturbative treatment.
Studies of other systems, notably Burgers equation and the Kraichnan
model, have been more successful \cite{falkovich2001}, providing insight 
into the
breakdown of self-similarity. 

In a recent paper \cite{crz2005} we added to
this small class of analytically tractable turbulent systems by showing
breakdown of self-similarity in a system of 
diffusing, aggregating masses with  a steady source (see
section~\ref{sec-model} for definition).
In this paper we provide a detailed presentation of the results of 
\cite{crz2005} and the methods developed to analyse this problem. We also
include several extensions of the ideas. These include the statistics of
one-point moments of the mass distribution which can be shown to exhibit
strong,  but in a sense trivial, anomalous scaling and a section which
establishes a heuristic connection between statistical field theory and
an effective kinetic theory description of aggregating particles (Smoluchowski
approximation).  We make extensive use of statistical field
theoretic techniques which, while well known in the context of equilibrium
condensed matter physics, we felt would benefit from some further 
explanation in the context of turbulent systems. The main thrust of the
calculations are explained in the main body of the article. Several appendices
are provided which discuss technical details and aspects specific to the 
aggregation model which we study.

The layout of the article is as follows. In section \ref{sec-model} we
define the ``mass model'' which underpins our study of aggregation. Next, in 
section \ref{sec-dimensionalAnalysis} we begin
where every study of a turbulent system should begin, namely with a simple
investigation of the implications of dimensional analysis. In this section
we explain the K41 phenomenology in the context of aggregation and derive
the scaling properties for the mass statistics which follow from the
assumption of self similarity. We then move on to the technical aspects of
the article. In section \ref{sec-fieldTheory} we give a detailed explanation
of how the mass model can be described in terms of an effective field 
theory using Doi's formalism. This  can, in turn, be mapped into a variant
of the $A+A\to A$ reaction diffusion model, a model which has been extensively
studied. In section \ref{sec-numerics} we present results from 
numerical simulations and exact computations of the scaling exponents of
the first and second order correlation functions. These confirm the 
breakdown of self-similarity in the mass model. The following two sections
contain a detailed exploration of this breakdown using 
renormalisation group arguments. Section \ref{sec-pertExpansion} outlines
the structure of the perturbation theory and Feynman rules for the
effective field theory and outlines how to perform the computations of
the relevant diagrams as expansions in $\epsilon=2-d$. In section \ref{sec-RG} 
we use reasonably standard RG
arguments to derive the Callan-Symanzik equation for the n-point correlation
functions. We solve these equations in $d<2$ to calculate the anomalous
scaling curve for the correlation functions as a first order epsilon-expansion.
We also solve the RG equations in $d=2$ to obtain the anomalous logarithmic
corrections to the mean field answers which one expects to find at the critical
dimension. In section \ref{sec-RSE} we return to a more physically-motivated
discussion and outline the connection between our renormalised perturbation
theory and an effective kinetic theory description based on the Smoluchowski
approximation.  We conclude with a brief summary of the main results and their
importance. Three appendices are provided which
supplement the main text with further technical explanations necessary for
a complete understanding of the work. With the inclusion of the appendices
we hope that the article becomes completely self contained.

\section{Definition of the Model}
\label{sec-model}
Consider a cubic lattice in $d$ dimensions whose lattice sites are occupied
by particles of integer masses. 
Multiple occupancy of a given site is permitted. 
Let the number of particles of mass $m$ on site $\V{x}$ at time $t$ be
denoted as 
$N_t(\V{x},m)$. At a given moment of time a configuration of the system is
determined by specifying the set of occupation numbers, 
$\{N_t(\V{x}_i,m)\}_{\V{x}_i \in {\mathbf R}^d}^{m\in {\mathbf Z}^+}$. These
configurations change in time due to three processes: diffusion, aggregation
and input. These processes are described below.
\begin{itemize}
\item{\bf Diffusion}
\begin{eqnarray*}
N_t(\V{x},m) &\to& N_t(\V{x},m) - 1,\\
N_t(\V{x}+\V{n},m) &\to& N_t(\V{x}+\V{n},m) + 1,
\end{eqnarray*}
where $\V{x}+\V{n}$ denotes a nearest neighbour of site $\V{x}$. This rule
describes the diffusive hopping of a particle at site $\V{x}$ to
the neighbouring site $\V{x}+\V{n}$. The rate at which such hopping occurs is 
$D N_t(\V{x},m) / (2 d)$. Here  $D$ is the diffusion
constant, assumed independent of the mass, and $2 d$ is the coordination number
of a cubic lattice in $d$ dimensions.
\item{\bf Aggregation}
\begin{eqnarray*}
N_t(\V{x},m_1) &\to& N_t(\V{x},m_1) - 1,\\
N_t(\V{x},m_2) &\to& N_t(\V{x},m_2) - 1,\\
N_t(\V{x},m_1+m_2) &\to& N_t(\V{x},m_1+m_2) + 1.
\end{eqnarray*}
This rule describes the aggregation of two particles of masses $m_1$ and $m_2$
at the same site $\V{x}$ to form a single new particle of mass
$m_1+m_2$ at site $\V{x}$.  The rate at which such aggregation events  occur is
$\lambda N_t(\V{x},m_1) N_t(\V{x},m_2)$ where $\lambda$ is the aggregation
rate, also assumed independent of mass.
\item{\bf Injection}
\begin{eqnarray*}
N_t(\V{x},m_0) &\to& N_t(\V{x},m_0) + 1.
\end{eqnarray*}
This rule describes the injection of new particles of mass $m_0$ at random 
sites throughout the system. We take $m_0=1$. The rate of injection
of new particles is $J$.
\end{itemize}
We will call the above model the mass model (MM). The MM has
3 physical parameters: the diffusion constant $D$, the aggregation rate 
$\lambda$, and input mass flux $J$.  In addition there
are two parameters associated with the lattice - the lattice spacing, $d x$,
and the smallest mass, $m_0$.

Various aspects of this model has been studied in different contexts. The
model with parallel update rules was initially studied as a simple model for
river networks \cite{scheidegger1967}. A comparison of different physical
quantities in this model relevant for river networks can be found in
\cite{dodds1999}. The solution to the single site mass distribution in one
dimension can be found in \cite{takayasu1989,huber1991}. From the exact
solution of the two point correlation function, the exponents governing the
single site mass distribution was found for arbitrary dimensions
\cite{RM2000}. The single site mass distribution was solved for mass
dependent aggregation kernel using Zakharov transformation \cite{CRZ1} and
using the Smoluchowski approximation 
in the context of submonolayer epitaxial thin film growth \cite{KMR1999}.
Some other contexts in which the model has been studied include granular bead
packs \cite{coppersmith1996}, nonequilibrium phase transitions
\cite{MKB2000,rajesh2004} and self organized criticality \cite{dhar1989}. All
these studies focused on the average mass distribution and its behaviour for
large masses. In this paper, we will be focussing mainly on multi-point correlation
functions.

\section{Dimensional Analysis and Self-Similarity Conjectures}
\label{sec-dimensionalAnalysis}
\subsection{Dimensional Considerations for the Mass Spectrum}

Before doing detailed calculations, we begin by asking what
we can learn about possible stationary states of the mass model from simple
dimensional considerations. The first quantity of interest in characterising
the long time behaviour of the model is the stationary mass spectrum,
$\langle N_m \rangle$.  $\langle N_m \rangle$ is the number of particles of 
mass $m$ per unit volume in the stationary 
state. Specifically, we would like to know how $\langle N_m \rangle$ 
scales with $m$ for
large values of $m$. Since we are hoping for universal behaviour in the limit 
of large $m$, we assume that the mass spectrum does not depend on the position
of the source, $m_0$. This assumption must be verified at
a later stage. The remaining dimensional parameters upon which $N_m$ could,
in principle, depend are $J$, $D$, $\lambda$ and , of course, $m$. We shall
perform most of our computations in units where $D=1$ but for the purposes
of dimensional analysis we shall keep $D$.

The dimension of $N_m$ is $[N_m] = {\rm M}^{-1} {\rm L}^{-d}$. The dimensions
of the other parameters are: 
$[J] = {\rm M} {\rm L}^{-d} {\rm T}^{-1}$, $[D] = {\rm L}^{2} {\rm T}^{-1}$ and 
$[\lambda] = {\rm L}^{d} {\rm T}^{-1}$. It is immediately evident that there
are too many dimensional parameters in the model to uniquely determine the 
mass spectrum. One can readily verify that for any scaling exponent, $x$,
and dimensionless constant, $c_1$, the formula
\begin{equation}
\langle N_m \rangle = c_1\, J^{x-1} D^\frac{(3-2x)d}{d-2}
\lambda^\frac{(d+2)x-2d-2}{d-2}m^{-x},
\label{eq-generalDimAnalysis}
\end{equation}
is a dimensionally correct expression for $\langle N_m \rangle$ 
which scales as $m^{-x}$. 
This is different to the dimensional argument used by Kolmogorov in his theory
of hydrodynamic turbulence. For that system, there is a unique dimensionally
correct combination of parameters giving the energy spectrum.
Eq. (\ref{eq-generalDimAnalysis}) allows us to pick out the scaling 
exponent, $x$, for the reaction and diffusion limited regimes:
\begin{eqnarray}
\label{eq-KZExponent}x^{\rm KZ} &=&\frac{3}{2},\\
\label{eq-K41Exponent}x^{\rm K41}&=&\frac{2d+2}{d+2}.
\end{eqnarray}
The above two exponents correspond to a different balance 
of physical processes in order to realise a stationary state. We briefly 
discuss each to explain the choice of nomenclature.
\begin{itemize}
\item
We shall call $x^{\rm KZ}$ the Kolmogorov--Zakharov (K--Z) exponent since it 
is the analogue for aggregating particles of the K--Z spectrum of wave
turbulence \cite{zakharovBook}
in the sense that it is obtained as the stationary solution of a 
mean field kinetic equation. This spectrum describes a reaction limited regime 
where diffusion plays no role.
\item
We shall call $x^{\rm K41}$ the Kolmogorov 41 (K41) exponent since it is 
a closer analogue of the $5/3$ spectrum of hydrodynamic turbulence
originally proposed by Kolmogorov in his 1941 papers on self-similarity in
turbulence. This is because in the Navier-Stokes equations there is no 
dimensional parameter like the reaction rate controlling the strength of the 
nonlinear interactions. This exponent describes a diffusion limited regime 
where the reaction rate, $\lambda$, plays no role, reactions being effectively 
instantaneous.
\end{itemize}

The case $x=1$ corresponds to one in which $\langle N_m \rangle$ does not
depend on the mass flux $J$. However, this is not of physical interest for
this problem.
On the other hand, each of the regimes
characterised by $x^{\rm KZ}$ and $x^{K41}$ carry a mass flux and is
relevant.

\subsection{Self-Similarity Conjectures and Multipoint Correlation Functions}

We are interested in more than just the average mass density $\langle N_m
\rangle$.  
To characterise correlations in the the mass model we must also
consider multipoint structure functions. Let 
$C_{n}(m_1,\ldots,m_n) (\Delta V)^n \prod_i dm_i$ be the probability of 
having particles of masses $m_i$ in the intervals $[m_i, m_i+dm_i]$ in a  
volume $\Delta V$ for $i=1\ldots n$.  $C_1(m)$ is the average mass density
$\langle N_m \rangle$. We ask 
how $C_n(m_1,\ldots,m_n)$ varies with mass when $m_1,\ldots,m_n \gg m_0$. 
In particular what is the value of the homogeneity exponent $\gamma_n$ 
defined through 
$C_n(\Gamma m_1,\ldots,\Gamma m_n) =\Gamma^{-\gamma_n} C_n(m_1,\ldots,m_n)$?

As for the density, dimensional analysis alone is insufficient. The formula
\begin{eqnarray}
\nonumber C_{n}(m_1,\ldots,m_n) &=& c_n\,J^{\gamma_n-n} 
D^\frac{(3n-2\gamma_n)d}{d-2} \lambda^\frac{(d+2)\gamma_n+(2d+2)n}{d-2}\\
\label{eq-CnDimAnalysis} & &\ \ \ \  (m_1\ldots m_n)^{-\frac{\gamma_n}{n}},
\end{eqnarray}
is dimensionally consistent for any exponent $\gamma_n$ where $c_n$ is
a dimensionless constant.  We are again assuming that the large mass behaviour
of the $C_n$'s is independent of $m_0$. The simplest way to obtain a 
theoretical prediction for the mass dependence of the $C_n$'s is to use a 
self-similarity conjecture similar to Kolmogorov's 1941 conjecture about the 
statistics of velocity increments in hydrodynamic turbulence. Assume that 
$C_n$ depends only on the masses $m_i$, mass flux $J$ and either the reaction 
rate, $\lambda$ or the diffusion coefficient $D$. Depending on which 
assumption we make, dimensional analysis allows us to determine the mass 
dimension of $C_n$. For the reaction limited case we obtain
\begin{equation}
\gamma_n^{\rm KZ} = \frac{3}{2} n,
\end{equation}
and for the diffusion limited case,
\begin{equation}
\gamma_n^{\rm K41} = \frac{2d+2}{d+2} n.
\label{kolmogorovanswer}
\end{equation}
Note that in both cases, the dependence of $\gamma_{n}$ on the index
$n$ is linear, reflecting the assumed self-similarity of the
statistics of the local mass distribution.
When $n=1$, $\gamma_{1}^{\rm K41} =(2 d+2)/(d+2)$ coincides with the result of
an exact computation of $\gamma_1$ for $d<2$ \cite{RM2000} so we expect that
the K41 conjecture is the appropriate theory in $d<2$. In $d>2$ it is known
that $\gamma_1=3/2$, hence 
$\gamma_{1}^{\rm KZ} = 3/2$ is the correct scaling for the density.
Therefore the KZ conjecture is appropriate in higher dimensions. For $d>2$,
the statistics of the MM should be accurately
described by mean field theory (KZ) and the self-similarity conjecture
should hold. In this paper we will concern ourselves with $d\leq 2$. 

The K41 self-similarity conjecture assumes that $C_n$ does not depend on 
$\lambda$, $m_0$, the lattice spacing, and the box size 
$\Delta V dm_1\ldots dm_n$.  The lack of dependence on the lattice spacing
is expected due to the renormalizability of the effective field theory 
describing the MM below two dimensions. We will however find an anomalous 
dependence of correlation functions on a length scale depending on the other 
parameters and the box size which leads to a violation of self-similarity.

\section{Field Theoretic Description of the Mass Model}
\label{sec-fieldTheory}
\subsection{Motivation}
\label{subsec-FT-motivation}
In order to check the validity of Kolmogorov conjecture for
the mass  model for $n>1$, we need to go beyond dimensional analysis.
We shall do this by constructing an effective
field theory that provides us with a continuum description of
the model. We are helped by the fact that the effective field theory
which describes the mass model is closely related to a much simpler
and well studied theory which describes the $A+A\to A$ reaction-diffusion
model. We shall then use standard techniques of statistical
field theory to extract information about the mass model from this
theory. As is well known, this model has critical dimension 2. In dimensions
greater than 2, a mean field description, characterised by K--Z scaling
works. In $d\leq 2$, this mean field description breaks down and correlations
between particles become important. We shall show how to take into account
these correlations and we demonstrate how they lead to the breakdown of
self-similarity in $d\leq 2$ by calculating the scaling behaviour of the
correlation functions, $C_n$, for large masses.  Unlike for the case 
hydrodynamic turbulence, such an analysis is possible for the mass model. This
is because in $d\leq 2$, the large mass statistics of the model in are 
governed by a perturbative fixed point of the renormalization
group flow in the space of coupling constants of the model. The order of this 
fixed point is $\ep=2-d$, which
allows one to compute the relevant scaling exponents in the form of
an $\ep$-expansion.

\subsection{Effective Action for the Mass Model}
\label{subsec-FT-effectiveAction}
Using Doi's formalism, it is possible to construct an effective
field theory of the mass model. The steps in the procedure are as
follows: 
\begin{enumerate}
\item
Write a {\em master equation} for time evolution of 
${\mathbf P}(\{N_t(\V{x}_i,m)\})$, the probability of finding the 
system in a given configuration,
$\{N_t(\V{x}_i,m)\}$. The master equation is linear and first order in time. 
\item
Introduce creation and annihilation operators, $a_{i,m}$ and  $a_{i,m}^\dagger$,
which create and destroy particles of mass $m$ at site $\V{x}_i$. Then convert 
the master equation into a Schroedinger equation :
\begin{displaymath}
\dd{ }{t}\, \left|\psi(t)\right> = -H[a_{i,m},a_{i,m}^\dagger]\, 
\left|\psi(t)\right>,
\end{displaymath}
using Doi's formalism (second quantisation) \cite{doi1976a,doi1976b}.
\item
Use the Feynman trick to derive a functional  integral measure which converts 
the second quantised Schroedinger equation into a continuous field theory.
\end{enumerate}
These steps are well described in the context of reaction-diffusion models
in \cite{cardyhttp,lee1994}. For the model of interest here, the procedure
is similar to the reaction-diffusion case with the algebra made slightly
more complicated by the necessity to keep track of a mass index for each 
particle. Explicit formulae for the master equation and Hamiltonian operator
of the MM are given in appendix \ref{app-MMHamiltonian}.

After going over to a path integral formulation of the Schroedinger equation
we can express the average density and other correlation functions as 
path integrals. Further detail on this procedure can  be found in appendix
\ref{sec-shiftExplanation}. For example the average density,
$N_t(m) = \langle \phi_{\V{x},m}(\tau)\rangle$, is given, in the notation
of appendix \ref{sec-shiftExplanation},  by 
\begin{equation}
\label{eq-TMFieldTheory}
N_t(m) = \int {\mathcal D}\phi {\mathcal D}\phi^* \phi_{\V{x},m}(\tau)\,
e^{-S_{\rm MM}[\phi,\phi^*, D,J,t,\lambda]},
\end{equation}
where
\begin{eqnarray}
\nonumber S_{\rm MM}[\phi, \phi^*, D,J,t,\lambda] &=& \int_0^t d\tau 
\int d^d\V{x}\,dm\ 
\left\{\phi^*\partial_t\phi \right. \\
& & \hspace{2.0cm} \left.+ H[\phi, \phi^*]\right\},
\label{eq-SMM}
\end{eqnarray}
and 
\begin{eqnarray*}
H[\phi, \phi^*] &=& D\,\nabla_{\V{x}}\phi_m^*\cdot\nabla_{\V{x}}\phi_m + \frac{J}{m}\delta(m-m_0)\phi_m^*\\
&-&\lambda\int dm_1dm_2 \left\{\delta(m_2-m-m_1)\right.\\
& &\left.\left[\phi^*_{m_2} - 2\phi^*_{m} - \phi^*_{m}\phi^*_{m_1}\right] 
\phi_{m}\phi_{m_1}\right\}.
\end{eqnarray*}

\subsection{Dimensional Analysis of the Effective Action}

As for the case of stochastic aggregation without source \cite{oleg2001}, it
is helpful to nondimensionalise the fields $\phi$ and $\phi^*$ and
express the action in eq. (\ref{eq-SMM}) 
solely in terms of dimensionless quantities.
Introduce dimensionless fields, $\bar{\phi}$ and $\bar{\phi}^*$, in eq. 
(\ref{eq-TMFieldTheory}) by the following rescalings: 
\begin{eqnarray*}
\tau &\to& t\, \tau,\\
\V{x} &\to& \sqrt{D t}\, \V{x},\\
m &\to& \lambda J t^2\, m,\\
\phi&\to& \frac{1}{\lambda^2 J t^3}\, \bar{\phi},\\
\phi^*&\to& \bar{\phi}^*,
\end{eqnarray*}
to obtain
\begin{equation}
\label{eq-TMdimensionless}
N_t(m) = \int {\mathcal D}\phi {\mathcal D}\phi^* \phi_{\V{x},m}(\tau)\, 
e^{-\frac{1}{g}S_{\rm MM}[\bar{\phi},\bar{\phi}^*, 1,1,1,1]},
\end{equation}
where the dimensionless interaction coefficient is
\begin{equation}
\label{eq-g}
g = D^\frac{d}{2}\, t^\frac{d-2}{2}\, \lambda.
\end{equation}
The fact that $g\to 0$ as $t\to\infty$ for dimensions greater than 2 and
$g\to\infty$ as $t\to\infty$ for dimensions less than 2 expresses the well
known fact that the critical dimension of the mass model is 2. We shall have
much more to say about this later.

\subsection{The Stochastic Smoluchowski Equation}
\label{subsec-FT-SSE}
It is possible to establish an exact map between the field theory 
in eq.~(\ref{eq-TMFieldTheory}) and the following 
stochastic integro-differential equation, \cite{lee1994, oleg2001}:
\begin{eqnarray} 
&& \left(\frac{\partial}{\partial t} -D \nabla^2 \right)
\phi(m) =
 \la \int_{0}^{m}  dm' \phi(m') \phi(m-m')
\nonumber\\
&&- 2\la \phi(m) N +\frac{J}{m_{0}}\delta (m-m_{0})
+i\sqrt{2\lambda}\phi(m) \eta(\xv,t), \label{sse}
\end{eqnarray}
where $N=\int_{0}^{\infty} dm'\phi(m')$,
$i=\sqrt{-1}$, and $\eta(\xv,t)$ is white noise in space and time:
\begin{equation}
\langle \eta(\xv,t) \eta(\xv',t')\rangle =\delta(t-t')\delta^d (\xv-\xv') .
\end{equation}
The technical details of this mapping can be found in appendix 
\ref{sec-shiftExplanation}.
Without the noise term, one recognises eq. (\ref{sse}) as the mean field
(Smoluchowski) equation of the model. Thus, all fluctuation
effects are encoded in the imaginary multiplicative noise term.

\subsection{Correspondence with $A+A \to A$ Model}
\label{subsec-FT-2A2A}
Eq. (\ref{sse}) simplifies after taking Laplace transform with
respect to the mass variable \cite{oleg2001}. Let
\begin{equation}
 R_{\mu} (\xv,t)=\int_{0}^{\infty} \!\!\! dm \phi(\xv,m,t)-\int_{0}^{\infty}
\!\!\!  dm \phi(\xv,m,t) e^{-\mu m}.
 \label{eq-defnR_mu}
\end{equation}
 Then,
 \begin{equation}
\left(\!\frac{\partial}{\partial t}\! -\! D \nabla^2\! \right)\!
R_{\mu}(\xv ,t) = -\la R_{\mu}^2+\frac{j_\mu}{m_{0}}+ 2 i
\sqrt{\la}R_{\mu}(\xv,t)\eta(\xv,t),
 \label{sre}
 \end{equation}
 where $j_\mu=J(1-e^{-\mu m_{0}})$.
In terms of field $R_{\mu}(\xv,t)$, eq. (\ref{sre}) becomes a stochastic
version of the rate equation for the $A+A \rightarrow A$
reaction in the presence of a source. Hence, the computation of
the average mass distribution in the mass  model reduces to solving a
one-species particle system with a $\mu$-dependent source and then
computing the inverse Laplace transform with respect to $\mu$. 
For example, to compute the average density, $\langle N_m(t) \rangle$, 
for the mass model, 
we first calculate $\langle R_\mu\rangle$, the average of the
solution of eq. (\ref{sre}) with respect to the noise, $\eta(\xv,t)$.
We then take the inverse Laplace transform with respect to $\mu$ and obtain
the density from eq. (\ref{eq-defnR_mu}). By applying the 
Martin--Siggia--Rose (MSR) procedure \cite{MSR}
to eq. (\ref{sre}), we can write 
$\langle R_\mu\rangle$ as a functional integral :
\begin{equation}
\langle R_\mu\rangle = \int {\mathcal D}R_\mu {\mathcal D}\widetilde{R}_\mu\ 
R_\mu\, e^{-S_{\rm RD}[R_\mu,\widetilde{R}_\mu]},
\end{equation}
where the effective action for the reaction-diffusion system described by
eq. (\ref{sre}) is
\begin{eqnarray}
\nonumber S_{\rm RD}[R_\mu,\widetilde{R}_\mu] &=& \int d\V{x}dt 
\left[ \vphantom{ \frac{j}{m_0}} \widetilde{R}_\mu(\partial_t-D\Delta)R_\mu 
+ \lambda \widetilde{R}_\mu R_\mu^2 \right.\\
& & \left. + \lambda \widetilde{R}_\mu^{2}R_\mu^2 - 
\frac{j}{m_0}\widetilde{R}_\mu \right].
\label{eq-Srd}
\end{eqnarray}
In order to compute higher order
correlation functions $C_{n}(m,t)$, we need to know
correlation functions of the form $\langle R_{\mu_{1}}(\xv,t)
R_{\mu_{2}}(\xv,t) \ldots R_{\mu_{n}}(\xv,t)\rangle$. These are
non-trivial, as the stochastic fields $R_{\mu}(\xv,t)$'s are
correlated for different values of $\mu$ via the common noise term
in eq. \ref{sre}. To clarify what is meant by this, we apply the MSR procedure
to two copies of eq. (\ref{sre}) describing the evolution of $R_{\mu_1}$
and $R_{\mu_2}$ respectively to obtain a functional integral representation
for $\langle R_{\mu_1} R_{\mu_2}\rangle$. This gives : 
\begin{eqnarray}
\nonumber \langle R_{\mu_1} R_{\mu_2} \rangle &=& \int {\mathcal D}R_{\mu_1} {\mathcal D}
\widetilde{R}_{\mu_1}{\mathcal D}R_{\mu_2} {\mathcal D}
\widetilde{R}_{\mu_2}\ R_{\mu_1} R_{\mu_2} \\
\nonumber & &\times e^{-S_{\rm RD}[R_{\mu_1},\widetilde{R}_{\mu_1}]}\, e^{-S_{\rm RD}[R_{\mu_2},\widetilde{R}_{\mu_2}]}\\
\label{eq-2pointAction}& &\times e^{-\int d\V{x}dt\ 2\lambda  
\widetilde{R}_{\mu_1} \widetilde{R}_{\mu_2} R_{\mu_1} R_{\mu_2}}.
\end{eqnarray}
The point to note here is that the path integral measure for correlations
of $R_\mu$ for {\em different} values of $\mu$ does not factorise owing
to the presence of the last term in eq. (\ref{eq-2pointAction}). Thus,
to compute $n$-point correlation
functions in the MM, one needs to analyse a system of $n$
stochastic rate equations of $A+A\rightarrow A$ theory coupled via
common noise terms of the form shown in eq. (\ref{eq-2pointAction}). Some
detailed explanation of the physical interpretation of higher order 
correlation functions in terms of the probability of multi-particle 
configurations in the particle system is provided in appendix \ref{app-SSE}.

\subsection{Feynman Rules}
\label{subsec-FT-FeynmanRules}
\begin{figure}
\begin{center}
\includegraphics[width=7.0cm]{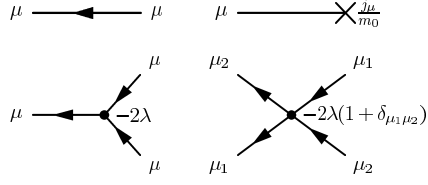}
\end{center}
\caption{\label{fig1} Propagators and vertices of the theory.}
\end{figure}

Solving eq. (\ref{sre}) perturbatively in $\lambda$ and $j$,
and then averaging over noise, one can derive the set of Feynman
rules for the computation of correlation functions.
Alternatively they can be written down 
directly from the action, eq. (\ref{eq-Srd}). See \cite{cardyhttp} for the
details of the procedure.  However care must be taken to
include the ``extra'' vertex which arises when computing correlations 
between fields with different $\mu$ indices.
The Feynman rules are summarised in  fig. \ref{fig1} with time increasing
from right to left. The slightly more
complicated prefactor for the quartic vertex takes into account the 
aforementioned ``extra'' vertex.

The propagator is just the regular diffusive Green's function which, in $d$
spatial dimensions, is 
\begin{equation}
G_0(\V{x}_2-\V{x}_1, t_2-t_1) =  4\pi(t_2-t_1)^{-\frac{d}{2}}\ 
e^{-\frac{\left|(\V{x}_2-\V{x}_1 \right|}{4(t_2-t_1)}},
\end{equation}
or
\begin{equation}
\label{eq-diffGreenFn}
\hat{G}_0(\V{k},t) = (2\pi)^{-\frac{d}{2}}\ e^{-k^2 t},
\end{equation}
in the momentum-time representation usually used in computations.

The one-point function, $\langle R_\mu\rangle$ is then given by the sum of
all diagrams constructed from the building blocks shown in fig. \ref{fig1} 
with a single outgoing line. 
Likewise, the $n$-point correlation function $\langle
R_{\mu_{1}}(\xv_{1},t_{1})$ $R_{\mu_{2}}(\xv_{2},t_{2})$ $\ldots
R_{\mu_{n}}(\xv_{n},t_{n})\rangle$ is given by the sum of the contributions of
all  diagrams which have $n$-outgoing lines. In section 
\ref{sec-pertExpansion} we shall turn to actual computations.

%

\section{Numerical Simulations and Multiscaling}
\label{sec-numerics}
\subsection{Numerical Simulations of the Mass Model in d=1}
\label{sec-numericalResults}
\begin{figure}
\includegraphics[width=12.0cm]{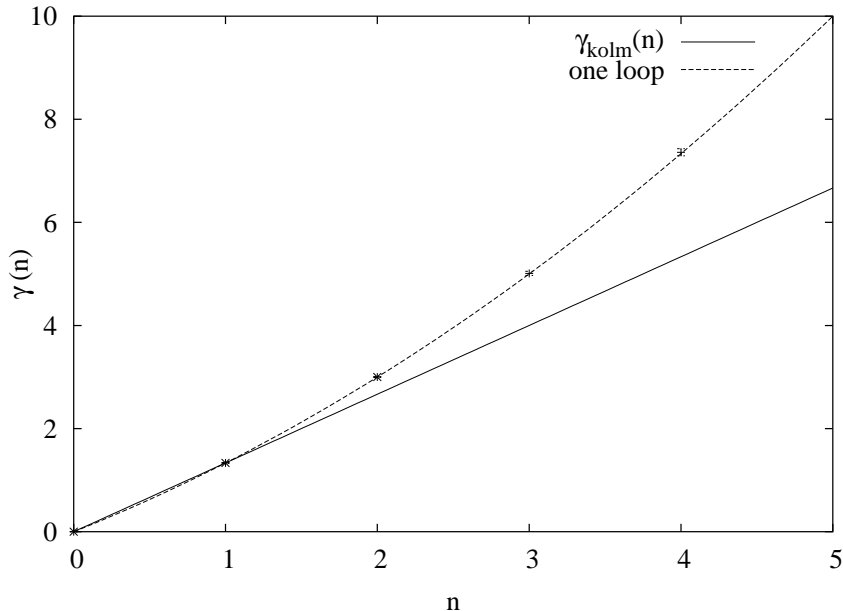}
\caption{\label{fig-numerics} $\gamma_n$ as a function of $n$ is one
dimension. The straight line shows the Kolmogorov answer
[eq. (\ref{kolmogorovanswer})]. The dotted line shows 
eq. (\ref{finalanswer}) 
with $\epsilon=1$ and terms of
order $\epsilon^2$ and higher set to zero. The values $\gamma_0$, $\gamma_1$ and
$\gamma_2$ are exact. $\gamma_3$ and $\gamma_4$ were obtained by Monte Carlo
simulations performed on a lattice of size $10^5$ and averaged over $2\times
10^7$ Monte Carlo time steps with $J=4 D$.
}
\end{figure}

We first look
at the results of Monte Carlo simulations of the MM which
confirm that there is indeed some interesting behaviour which requires
explanation. In particular, numerical simulations show a breakdown 
of self-similarity in the mass model in one dimension and 
multiscaling of the correlation functions, $C_n$.
The results are shown in fig. \ref{fig-numerics}. 

\subsection{Constant Flux Relation - Analytic Confirmation of Multiscaling}

We know that the K41 hypothesis works for $n=1$. From
fig. \ref{fig-numerics}, it is clear that the Kolmogorov scaling breaks
down for $n>1$.
It is also possible to analytically confirm that $\gamma_n\neq 
\gamma^{K41}_n$
in $d<2$ by computing $\gamma_2$. From the definition of $\gamma_2$, if
follows that
\begin{eqnarray}
\Phi_{2}(m_{1}, m_{2})=\bigg(
\frac{1}{m_{1}m_{2}}\bigg)^{\gamma_2/2} \phi\bigg(
\frac{m_{1}}{m_{2}}\bigg),\label{sc2}
\end{eqnarray}
where $\phi$ is an unknown scaling function which satisfies $\phi (x)= \phi
(1/x)$ due to a symmetry. Our aim is to compute $\gamma_{2}$
without using the $\ep$-expansion which we shall use in section \ref{sec-RG}
to compute $\gamma_{n}$ for general $n$.

As we are interested in $\Phi_{2}(m_{1}, m_{2})$ for $m_{1},
m_{2}>0$,
\begin{eqnarray}
\Phi_{2}(m_{1}, m_{2})=\int_{\sigma-i\infty}^{\sigma+i\infty}
\int_{\sigma-i\infty}^{\sigma+i\infty} d\mu_{1} d\mu_{2} \langle
R_{\mu_{1}} R_{\mu_{2}} \rangle, e^{-m_1\mu_1}\,e^{-m_2\mu_2},
\end{eqnarray}
where $R_{\mu}$ solves eq. (\ref{sre}). Due to eq. (\ref{sc2}),
\begin{eqnarray}
\langle
R_{\mu_{1}} R_{\mu_{2}} \rangle=\bigg( \frac{1}{\mu_{1}\mu_{2}}
\bigg)^{1-\gamma_2/2}\psi\bigg( \frac{\mu_{1}}{\mu_{2}} \bigg), \label{sc2l} 
\end{eqnarray}
where $\psi$ is an unknown scaling function.
To find the large $m_{1}, m_{2}$ asymptotics of $\Phi$, we need to
know the small $\mu_{1}, \mu_{2}$ asymptotics of $\langle
R_{\mu_{1}} R_{\mu_{2}} \rangle$. Averaging  eq. (\ref{sre}),
with respect to noise and setting $\partial_{t} \langle R_{\mu} \rangle=0$
in the large time limit, we find that $\langle
 R_{\mu} R_{\mu} \rangle =\frac{j}{\lambda m_{0}} \approx
 \frac{J\mu}{\lambda}$ for $\mu \ll m_{0}$. Comparing this result with
eq. (\ref{sc2l})
we find that $\gamma_2=3$.

Note that $\gamma_2$ does not depend on dimension, $d$, of the lattice. 
Therefore, it is correctly predicted by mean field theory. The
non-renormalization of $\gamma_2$ by diffusive fluctuations
can be explained by mass conservation or, more precisely by
constancy of the average flux of mass in the mass space, see
\cite{CRZ1} for more details. Here, we simply wish to point out that the
exact answers for $\gamma_1$ and $\gamma_2$ establish
multiscaling non-perturbatively: the points $(0,0)$,
$(1,\gamma_1)$ and $(2, \gamma_2)$ do not lie on the same straight line.

Due to its close connection with mass conservation, the law
$\gamma_2=3$ is a counterpart of the $4/5$ law of
Navier-Stokes turbulence. Recall, that $4/5$ law states 
that the third order longitudinal structure function of the
velocity field scales in the inertial range as the first power of the
separation. It is interesting to notice, that Kolmogorov
theory respects $4/5$ law in Navier-Stokes turbulence, but
violates $\gamma_2=3$ in the MM.

\section{Perturbative Expansion for Correlation Functions}
\label{sec-pertExpansion}
\subsection{Mean Field Analysis}
\begin{figure}
\begin{center}
\includegraphics[width=7.0cm]{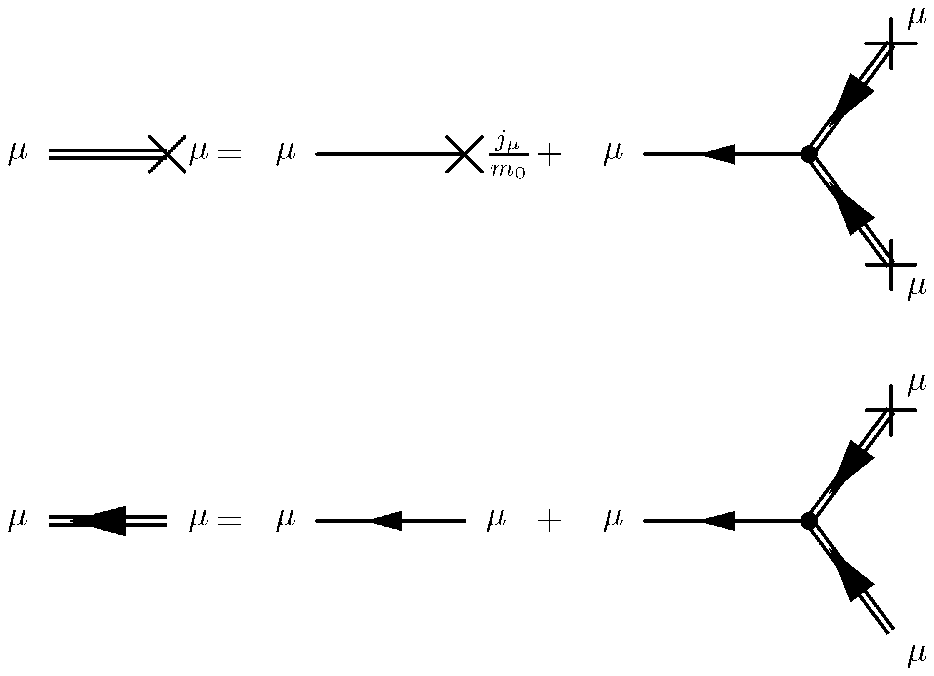}
\end{center}
\caption{\label{fig2} Diagrammatic form of mean field equations for
$R^{\rm mf}_\mu(t)$ and the tree-level propogator, $G^{\rm mf}_\mu({\bf 2}; {\bf 1})$.}
\end{figure}

The mean field theory associated with the field theory described by the
effective action, eq. (\ref{eq-SMM}) can be thought of in several complementary ways.
Let us suppose that the reaction rate, $\lambda$, is the smallest parameter
in the problem. This means that the dimensionless interaction coefficient, $g$,
given by eq. (\ref{eq-g}) is small. In this case, the path integral
in eq. (\ref{eq-TMdimensionless}) can computed in the limit $g\to 0$ 
using the saddle point method. In this limit, $\phi$ satisfies the
Euler-Lagrange equation (expressed in dimensional variables) :
\begin{eqnarray}
&& \left(\frac{\partial}{\partial t} -D \nabla^2 \right)
\phi(m) =
 \la \int_{0}^{m}  dm' \phi(m') \phi(m-m')
\nonumber\\
&&- 2\la \phi(m) N +\frac{J}{m_{0}}\delta (m-m_{0}), 
\label{eq-EL}
\end{eqnarray}
which we recognise as the mean field equation derived for classical aggregation
problems by Smoluchowski.  Now, if $g\to 0$ then it follows that the noise term
disappears from the non-dimensionalised version of the stochastic Smoluchowski
equation, eq. (\ref{sse}), leaving us with a deterministic equation for the
density, which is again the classical Smoluchowski equation. For readers 
interested in the analogy between stochastic aggregation and wave turbulence,
the mean field Smoluchowski equation is the analogue of the kinetic equation.

In \cite{CRZ1} we studied in great detail the stationary state of eq.
(\ref{eq-EL}) and showed that the spectrum
\begin{equation}
\label{eq-KZ}
N_m = \sqrt{\frac{J}{4\pi\lambda}}\, m^{-\frac{3}{2}},
\end{equation}
is the exact stationary solution as $t\to\infty$. This solution carries a 
constant flux, $J$, of mass from small masses to large. This is the Kolmogorov--
Zakharov spectrum of the mass model which we identified from dimensional
considerations in section \ref{sec-dimensionalAnalysis} as corresponding to
a reaction limited regime.

When do we expect the mean field answers to be correct? The K--Z solution is
established in the limit of large times. Since the mean field results become
exact in the limit $g\to 0$, eq. (\ref{eq-g}) implies that the spectrum
(\ref{eq-KZ}) should be correct for $d>2$. For $d\leq 2$ the mean field 
approximation quickly breaks down and we must take into account the effect of
fluctuations. This will be the main objective of the rest of this article.

Let us now  identify clearly the terms in the diagrammatic expansion which give the
mean field answers so that we can see how to use our formalism to compute the
fluctuations about the mean field. Since the mean field kinetic equation
corresponds to the deterministic limit of the stochastic Smoluchowski equation,
the corresponding field theory has no loops. Therefore we expect the
mean field answers for the average density to correspond to the sum of
all tree diagrams with a single outgoing line. Let us now analyse these.

Let $R_{mf}$, denoted by
a thick line with a cross, be the contribution to $R$ from all
tree level diagrams. The equation satisfied by $R_{mf}$ is shown
is diagrammatic form in fig. \ref{fig2}A. In equation form, it
reads
 \begin{equation}
 \frac{d R_{mf}}{dt} = \frac{j_\mu}{m_0} - \lambda R_{mf}^2.
 \end{equation}
 This is easily solved to give
 \begin{equation}
 R_{mf}(t)= \sqrt{\frac{j_\mu}{m_0 \lambda}}
\tanh\left( \sqrt{\frac{j_\mu \lambda}{m_0}} t \right)
\stackrel{t\rightarrow \infty}{\longrightarrow} \sqrt{\frac{j_\mu}{m_0
\lambda}}.
\end{equation}
Performing the inverse Laplace transform in the limit $\mu\to 0$ we find that 
as $m \to \infty$,
\begin{displaymath}
N_m \sim \sqrt{\frac{J}{4\pi\lambda}}\, m^{-\frac{3}{2}},
\end{displaymath}
and recover the K--Z spectrum as we should. Both the constant and the exponent
agree with those obtained by the Zakharov transformation of the mean field
kinetic equation confirming that our approach makes sense.

It is convenient to define $G^{\rm mf}_\mu(x_2 t_2;x_1 t_1)$ as the
propagator that includes all the tree level diagrams. The equation
obeyed by it is shown in fig. \ref{fig2}B. The solution is
\begin{eqnarray}
 G^{\rm mf}_\mu({\bf 2}; {\bf 1})&=&
 G_{0}({\bf 2}; {\bf 1})
\left[\frac {\cosh \sqrt{\frac{j \lambda}{m_0}} t_1 } {\cosh
\sqrt{\frac{j \lambda}{m_0}} t_2 }
\right]^2,\\
&\stackrel{t_{1,2}\rightarrow \infty}{\longrightarrow} &
 G_{0}({\bf 2}; {\bf 1}) e^{-\Omega (t_2-t_1)},
 \end{eqnarray}
where $G_{0}$ is the Green's function of the linear diffusion equation, 
eq. (\ref{eq-diffGreenFn}), and 
\begin{equation}
\Omega_\mu = 2\sqrt{\frac{j_\mu\lambda}{m_{0}}},
\end{equation}
 is the inverse of the mean field response time.

\subsection{Loop Expansion}
In order to take into account fluctuations about the mean field answer
we need to compute diagrams with loops. Ordering the terms in the perturbation
series according to the number of loops is known as a {\em loop expansion}.
Using the mean field density and response functions computed above simplifies 
the task of computing the sum of all diagrams with a given number of loops.
We now demonstrate by power counting that loop expansion of the mean mass
distribution corresponds  to weak coupling expansion with respect
to $\lambda$. The quantity $\langle R(\xv,t) \rangle $ is given by
the sum of all diagrams with one outgoing line built out of blocks
shown in fig. \ref{fig1}. Consider such a diagram containing $L$
loops, $V$ vertices and $N$ $R_{mf}$-lines. The corresponding
Feynman integral contains (in the mixed momentum-time
representation)  $d L$ momentum integrals and $V$ time integrals.
Hence, the integration over all times and momenta produces the
factor $\Omega^{-V+ d L/2}\sim \la^{-\frac{V}{2}+\frac{dL}{4}}$.
The $N$ $R_{mf}$-lines produce the factor $R_{mf}^{N}\sim
\la^{-N\la/2}$. A factor $\la^{V}$ comes from $V$ vertices of the
graph. Hence the corresponding Feynman integral is proportional to
$\la^{-\frac{N}{2} +\frac{V}{2}+ \frac{dL}{4}}$. Note also that
the number of triangular vertices in the graph is equal to $N-1$
and the number of quadratic vertices is equal to the number of
loops $L$. Thus the total number of vertices is given by
$V=L+N-1$. Therefore, any $L$-loop graph contributing to the
average mass distribution is proportional to
$\la^{-\frac{1}{2}+\frac{L}{2}(1+\frac{d}{2})}$. We conclude that
loop expansion corresponds to the perturbative expansion of $R$
around the mean field value with the parameter
$\la^{\frac{2+d}{4}}$.

\subsection{Breakdown of Loop Expansion} 
\label{sec-breakdownOfLoopExp}
The conditions under which the loop corrections to the mean field
answer can be neglected are most simply derived using dimensional
analysis. The scale of diffusive fluctuations is given by the only
constant of dimension length which can be constructed out of $\mu$
and $J$: $L_{D}= (\mu J)^{-1/(d+2)}$. The dimensionless expansion
parameter in the loop expansion above is $g_{0}(\mu)=\la
L_{D}^{\ep/2}$, where $\ep=2-d$. The large mass behaviour of $N_m$
is determined by the small-$\mu$ behaviour of $R_{\mu}$. In $d< 2$,
$g_{0}$ goes to infinity in the limit $\mu \rightarrow 0$ and the
loop expansion breaks down. Thus a re-summation of the loop expansion
is needed in order to extract the large-$m$ behaviour of $N_m$ in low
dimensions.

\subsection{Calculation of One Loop Corrections to Mean Field Theory}
\label{sec-oneLoopAnswers}
\begin{figure}
\begin{center}
\includegraphics[width=7.0cm]{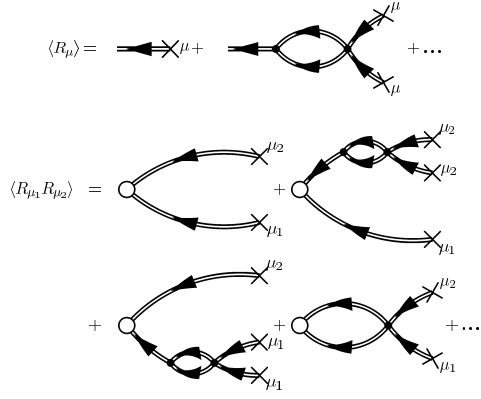}
\end{center}
\caption{\label{fig-oneLoop} Zeroth and first order terms in the loop expansions for $\langle R_\mu\rangle$ and $\langle R_{\mu_1} R_{\mu_2} \rangle$}
\end{figure}

Let us now compute $\langle R_\mu\rangle$ and
$\langle R_{\mu_1} R_{\mu_2} \rangle$ to one loop order. The 
diagrams are  shown in fig. \ref{fig-oneLoop}. The corresponding 
algebraic expressions are evaluated by dimensional regularisation in
dimension $d$, no longer necessarily an integer.

\begin{eqnarray}
\nonumber \langle R_\mu\rangle &=& R^{\rm mf}_\mu + R^{(1)}_\mu + \ldots\\
\label{eq-densityOneLoop}&=& \sqrt{\frac{j_\mu}{\lambda m_0}}
\left[1+\frac{\lambda^2 \Omega_\mu^\frac{d-4}{2}}{(4\pi)^\frac{d}{2}} 
\Gamma\left(\frac{\ep}{2}\right)  \sqrt{\frac{j_\mu}{\lambda m_0}} + 
\ldots\right],
\end{eqnarray}
where $\ldots$ represent terms of higher order in $\lambda$ which 
necessarily have more loops. In this formula we have introduced
the quantity $\epsilon$, defined as
\begin{equation}
\epsilon = 2-d,
\end{equation}
to measure the deviation of the dimension of the system from the
critical dimension.

The four diagrams for $\langle R_{\mu_1} R_{\mu_2} \rangle$,  shown 
in fig. \ref{fig-oneLoop}, give the following respective 
contributions : 
\begin{eqnarray}
\nonumber \langle R_{\mu_1}R_{\mu_2}\rangle &=& R^{\rm mf}_{\mu_1}
R^{\rm mf}_{\mu_2} - \frac{2 \lambda}{(8\pi)^\frac{d}{2}} 
\frac{R^{\rm mf}_{\mu_1}R^{\rm mf}_{\mu_2}}{(\Omega_{\mu_1}\!\!+\!
\Omega_{\mu_2})^\frac{\epsilon}{2}}\Gamma\left(\frac{\ep}{2}\right) \\
&&\label{eq-2pointOneLoop}
+ R^{(1)}_{\mu_1}R^{\rm mf}_{\mu_2} 
+ R^{\rm mf}_{\mu_1} R^{(1)}_{\mu_2}\ldots,\\
\nonumber &=& \langle R_{\mu_1}\rangle \langle R_{\mu_2}\rangle 
\left[1-\frac{2 \lambda}{(8\pi)^\frac{d}{2}} \frac{1}
{(\Omega_{\mu_1}\!\!+\!\Omega_{\mu_2})^\frac{\epsilon}{2}}
\Gamma\left(\frac{\ep}{2}\right)  + \ldots\right],
\end{eqnarray}
an expression which is correct to one loop order. Note that the
second diagram in the expression for 
$\langle R_{\mu_1} R_{\mu_2} \rangle$ describes the correlation between
$R_\mu$ fields for different values of $\mu$ and prevents the 
factorisation of the 2-point function into a product of 1-point 
functions.  We shall need these expressions again to when we use  RG 
to resum the loop expansion.

For a given mass scale, $m$, there is a corresponding $\mu$ scale, 
$1/m$ and a corresponding length scale, $L_\mu$ defined as
\begin{equation}
\label{eq-L}
L_\mu = \left(\frac{j_\mu}{D m_0}\right)^{-\frac{1}{d+2}}.
\end{equation}
At this point, let us also define the dimensionless reaction
rate, $g$, as 
\begin{equation}
\label{eq-dimensionlessg}
g = \lambda L_\mu^\epsilon.
\end{equation}
In what follows, it shall be convenient to express 
eq. (\ref{eq-densityOneLoop}) and eq. (\ref{eq-2pointOneLoop}) in terms 
of $L_\mu$ and $g$. This gives
\begin{eqnarray}
\label{eq-densityOneLoop2}\langle R_\mu\rangle = L_\mu^{\epsilon-2}\frac{1}
{\sqrt{g}} \left[1+\frac{1}{4(2\pi)^{1-\frac{\epsilon}{2}}} 
\Gamma\left(\frac{\ep}{2}\right)\, g^{1-\frac{\epsilon}{4}}   +
\ldots\right],
\end{eqnarray}  
and
\begin{eqnarray}
\label{eq-2pointOneLoop2}\langle R_{\mu_1}R_{\mu_2}\rangle &=& \langle R_{\mu_1}\rangle \langle R_{\mu_2}\rangle \left[1-\frac{g^{1-\frac{\epsilon}{4}}}{(4\pi)^{1-\frac{\epsilon}{2}}} \Gamma\left(\frac{\ep}{2}\right)\right.\\ 
\nonumber && \times \left. \left(\left(\frac{L_{\mu_1}}{L_\mu}\right)^{\frac{\epsilon-4}{2}}\!\!\!\!\!+\left(\frac{L_{\mu_2}}{L_\mu}\right)^{\frac{\epsilon-4}{2}}\right)^{-\frac{\epsilon}{2}}+\!\!\! \ldots\right].
\end{eqnarray}
Note that the $\mu$ dependence of this expression is illusory
since $g$ also depends on $\mu$.
To study the behaviour of the corrections to mean field
answers which we have just calculated, we need to study the large
$m$ behaviour of the Laplace transforms with respect to the $\mu$'s 
of the expressions in eq. (\ref{eq-densityOneLoop}) and eq.
(\ref{eq-2pointOneLoop}). Simple calculation shows that the
second terms inside the square brackets in these expressions 
diverge as the $\mu$'s are taken to zero when $\epsilon>0$ 
signifying a breakdown of the loop expansion. This is as  
expected from the power counting argument of section 
\ref{sec-breakdownOfLoopExp}.

\section{Renormalisation Group Analysis for $d<2$}
\label{sec-RG}
\subsection{Epsilon Expansion}

The loop expansions for correlation functions computed in
section \ref{sec-oneLoopAnswers} are expansions in powers of
the dimensionless reaction rate, $g$. The problem is that in 
$d<2$, these expansions become badly ordered as we approach 
$L_\mu \to \infty$.  However, since we 
have computed correlation functions in arbitrary dimension, $d$,
we can convert the loop expansions into expansions in
$\epsilon=2-d$ at a fixed value of $g$. For $\epsilon \ll 1$,
eq. (\ref{eq-densityOneLoop2}) and eq. (\ref{eq-2pointOneLoop2}) can 
be written :
\begin{eqnarray}
\label{eq-densityOneLoop3}\langle R_\mu\rangle = L_\mu^{\epsilon-2}\frac{1}
{\sqrt{g}} \left[1+\frac{g}{4\pi\epsilon} + \ldots\right],
\end{eqnarray}
and
\begin{eqnarray}
\label{eq-2pointOneLoop3}\langle R_{\mu_1}R_{\mu_2}\rangle &=& \langle R_{\mu_1}\rangle \langle R_{\mu_2}\rangle \left[1-\frac{g}{2\pi\epsilon} + \ldots\right].
\end{eqnarray}
Of course these series are still badly ordered as 
$\epsilon\to 0$. The idea
is to replace certain correlation functions with appropriate
renormalised quantities, also expressed as expansions in 
$\epsilon$, such that the renormalised counterparts of 
the above expressions are well-ordered in $\epsilon$. The 
final pay-off comes when we find that these expressions 
remain well ordered in $\epsilon$ even when we take the limit
$L_\mu\to \infty$ because of the presence of a {\em perturbative
fixed point}, a structural feature of the theory which we must 
now explain in order to make sense of this scheme.

\subsection{Renormalised Reaction Rate and $\beta$-function}
\label{sec-betaFunction}
\begin{figure}
\begin{center}
\includegraphics[width=7.0cm]{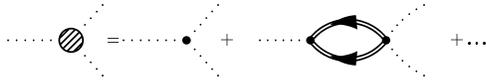}
\end{center}
\caption{\label{fig-lambdaR} Diagrams contributing to the renormalisation of
the reaction rate.}
\end{figure}

The presence of  a perturbative fixed point for the
$A+A\to A$ model was originally pointed out  by Peliti 
\cite{peliti}. The corresponding calculations in the presence
of a source were done by Droz\cite{droz}. This is sufficient 
to deal with the problem at hand. Nevertheless we shall 
paraphrase their arguments here for the sake of completeness.

Let us define a renormalised reaction rate, $\lambda_R$, as
the amputated 3-point vertex function shown diagrammatically in
fig. \ref{fig-lambdaR}. After performing the algebra we find
\begin{equation}
\lambda_R = \lambda \left[ 1 - \frac{1}{2(2\pi)^\frac{d}{2}}
\Gamma\left(\frac{\epsilon}{4}\right) g^{1-\frac{\epsilon}{4}} +
\ldots\right].
\end{equation}
Now we introduce a dimensionless renormalised reaction rate,
$g_R = \lambda_R L_\mu^\epsilon$, as we did in eq.
(\ref{eq-dimensionlessg}), which is given by
\begin{equation}
g_R = g   - \frac{1}{2(2\pi)^{1-\frac{\epsilon}{2}}}
\Gamma\left(\frac{\epsilon}{4}\right) g^{2-\frac{\epsilon}{4}} + \ldots.
\end{equation}
For small values of $\epsilon$ this can be written as
\begin{equation}
\label{eq-gRExpg}
g_R = g - g_*^{-1}(\epsilon) g^2 + \ldots,
\end{equation}
where
\begin{eqnarray}
\label{eq-gstar}
g_*^{-1}(\epsilon) &=&  \frac{1}{2(2\pi)^{1-\frac{\epsilon}{2}}}
\Gamma\left(\frac{\epsilon}{4}\right)\\
\nonumber  &=& \frac{1}{2\pi\epsilon}+o(1) \hspace{0.5cm}\mbox{$\epsilon\ll 
1$}.
\end{eqnarray}
Inverting eq. (\ref{eq-gRExpg}) allows us to convert 
perturbative expansions in the bare reaction rate, $g$, into 
expansions in the renormalised reaction rate, $g_R$. We find
\begin{equation}
\label{eq-gExpgR}
g = g_R + g_*^{-1}(\epsilon) g_R^2 + \ldots.
\end{equation}
The crucial point to all of this analysis is the following
observation. Although for positive $\epsilon$, $g$ diverges as 
$L_\mu\to\infty$, rendering perturbative expansions in $g$ 
useless for capturing the large mass behaviour of the theory, 
we will find that $g_R$ remains finite as $\mu\to\infty$. 
Furthermore, $g_R$ tends to a value which is of order 
$\epsilon$. Therefore for small $\epsilon$ we can use eq. 
(\ref{eq-gExpgR}) to convert expansions in $g$ into expansions 
in $g_R$ which then have a better chance of remaining 
non-singular when we take $L_\mu\to\infty$. The process of
replacing $g$ with $g_R$ is usually called {\em coupling
constant renormalisation} in the literature.

The large mass behaviour of the renormalised reaction rate is
determined by the $\beta$-function of the theory defined as
\begin{equation}
\label{defn-beta}
\beta(g_R) = \left.\left(L_\mu\pd{g_R}{L_\mu}\right)\right|_\lambda.
\end{equation}
Using the fact that $L_\mu\pd{ }{L_\mu} g^n = n\epsilon g^n$
together with eq. (\ref{eq-gRExpg}) and eq. (\ref{eq-gExpgR}) we
quickly find
\begin{equation}
\label{eq-beta}
\beta(g_R) = \epsilon g_R (1 - g_*^{-1}(\epsilon) g_R + \ldots).
\end{equation}
Eq. (\ref{defn-beta}) now tells us how $g_R$ changes as 
we vary $L_\mu$. Solving this differential equation with the initial
condition $g(L_0) = g_0$ determines how the reaction rate varies with scale.
The behaviour is different in $d=2$ and $d<2$. In $d<2$,
\begin{equation}
L_\mu\pd{g}{L_\mu} = \epsilon g (1-\frac{g}{2\pi\epsilon}),
\end{equation}
so that
\begin{equation}
g_R(L_\mu) = \frac{g_0L_\mu^\epsilon}{(1-\frac{g_0}{2\pi\epsilon})L_0^\epsilon+\frac{g_0}{2 \pi
\epsilon}L_\mu^\epsilon}.
\end{equation}
We note that $g_R$ goes to a fixed point value of $2\pi\epsilon$ (+corrections
of $O(\epsilon^2)$) as $L_\mu\to\infty$ irrespective of the initial values of
$g_0$ and $L_0$. For simplicity
we can take $L_0=(1-\frac{g_0}{2\pi\epsilon})^{-1}$ giving
\begin{equation}
\label{eq-gdlt2}
g_R(L_\mu) = \frac{g_0 L_\mu^\epsilon}{1+\frac{g_0}{2 \pi \epsilon}L_\mu^\epsilon}.
\end{equation}
This universal behaviour as $L_\mu\to\infty$ is what is meant when we say that 
the renormalisation group flow has a perturbative fixed point in $d<2$. 

In $d=2$, $\epsilon=0$ so that
\begin{equation}
L_\mu\pd{g}{L_\mu} = -\frac{g^2}{2\pi},
\end{equation}
which gives
\begin{equation}
\label{eq-gdeq2}
g_R(L_\mu) = \frac{g_0}{1+\frac{g_0}{2 \pi }\log
\left(\frac{L_\mu}{L_0}\right)}.
\end{equation}
Thus in $d=2$ the reaction rate decays to 0  as $L_\mu\to\infty$ but 
logarithmically slowly and, unlike in the case $d<2$, retains some memory of 
the small scale cut-off, $L_0$.

\subsection{Average density in $d<2$}
\label{sec-avgDensity}
Let us now show how all of this technology works by calculating
the large mass behaviour of the average density.  We define
the renormalised density, $\langle R_\mu\rangle_{\rm R}$, 
by using eq. (\ref{eq-gExpgR}) to replace $g$ with the
renormalised reaction rate, $g_R$ in eq. 
(\ref{eq-densityOneLoop3}). Using eq. (\ref{eq-gstar}), a
Taylor expansion shows that the replacement of $g$ with $g_R$
cancels the $\epsilon$-singular term so we are left with
\begin{eqnarray}
\label{eq-renormalisedR_mu}\langle R_\mu\rangle_{\rm R} = L_\mu^{\epsilon-2}\frac{1}{\sqrt{g_R}} \left[1+ o(g_R^2) \right].
\end{eqnarray}
Now as $L_\mu \to \infty$, $g_R \to g_*$ so we can now take
the limit to obtain
\begin{equation}
\langle R_\mu\rangle_{\rm R} \sim
L_\mu^{\epsilon-2}\frac{1}{\sqrt{2\pi\epsilon}} \left[1+ o(\epsilon^2)
\right].
\end{equation}
Finally note that as $\mu\to 0$, 
\begin{displaymath}
L_\mu \sim \left(\frac{J \mu}{m_0}\right)^{-\frac{1}{d+2}},
\end{displaymath}
allowing us to perform the inverse Laplace Transform
required to return to mass space. Using the definition, eq.
(\ref{eq-defnR_mu}), of $R_\mu$ we finally find
\begin{equation}
\label{eq-renormalisedDensity}
N_m \stackrel{\sim}{\scriptstyle m\to\infty} -\frac{1}{\Gamma\left(-d/d+2\right)}\, \frac{1}{\sqrt{2\pi\epsilon}} \left[1+ o(\epsilon^2) \right]\, m^{-\frac{2d+2}{d+2}},
\end{equation}
giving a scaling exponent which we know to be correct \cite{RM2000}.
We recognise this as the K41 exponent. As discussed in section
\ref{sec-dimensionalAnalysis} we could have obtained this answer simply
from dimensional arguments once we recognised that the
reaction rate is renormalised away to infinity and hence cannot
play any role in the answer. However this would ignore the
possibility of anomalous dimensions. By calculating the one
loop corrections to $\langle R_\mu\rangle$ we confirmed the
absence of any relevant (for $\langle R_\mu\rangle$) couplings 
other than the reaction rate itself. We shall find that this
is not the case for the higher order correlation functions.

\subsection{Higher order moments of the density}

A natural object to study to gather more information about the mass
distribution function would be moments of the density of the form 
$M_n(m) = \left<N_m^n\right>$. As explained in the appendix 
\ref{sec-shiftExplanation} (see eq. \ref{cor} and the explanation
thereafter) these moments exhibit ``extreme'' anomalous scaling 
characterised by Burgers-like scalings :
\begin{equation}
\label{eq-extremeAnomalousScaling}
\left< N_m^n \right> \sim \left< N_m\right>  \stackrel{\sim}{\scriptstyle m\to\infty}  m^{-\frac{2d+2}{d+2}}.
\end{equation}
For the MM however, this  anomaly is somewhat trivial from a 
physical perspective. It arises because large masses become large by absorbing
almost all nearby particles. Thus asymptotically, the number of heavy particles
on a given lattice site ends up being either zero or one. Taking moments of
such a distribution will always give the behaviour described by eq.
(\ref{eq-extremeAnomalousScaling}).  However, the analysis of appendix 
\ref{sec-shiftExplanation} which allows one to extract this essentially 
non-mean field behaviour from an initially weakly coupled theory is not trivial
and can be expected to yield interesting results in other contexts.
To observe true the multiscale structure of the mass model one should really 
study multipoint correlation functions. We do this next.

\subsection{Higher order multi-point correlation functions in $d<2$}

\begin{figure}
\begin{center}
\includegraphics[width=7.0cm]{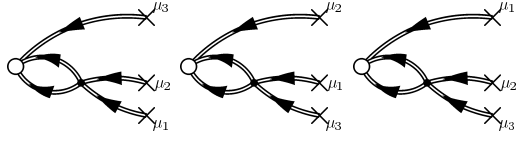}
\end{center}
\caption{\label{fig-3pointFn}  Diagrams contributing to the anomalous dimension of the 3-point function $\langle R_{\mu_1} R_{\mu_2} R_{\mu_3}\rangle$ at the one loop level.}

\end{figure}

The analysis for the higher order correlation
functions is not quite so simple as for the density.
By replacing $g$ with $g_R$ in eq. (\ref{eq-2pointOneLoop2})
we  get 
\begin{equation}
\langle R_{\mu_1}R_{\mu_2}\rangle_{g\to g_R} = \langle R_{\mu_1}\rangle_{\rm R} \langle R_{\mu_2}\rangle_{\rm R}  \left[1-\frac{g_R}{2\pi\epsilon} + o(g_R^2)\right].
\end{equation}
We see that we have removed the 
$\epsilon$-singularities from the $\langle R_\mu\rangle$ factors
but the singularity inside the square brackets remains. The
correct definition of the renormalised 2-point function must
include renormalisation of the amplitude of $C_2$, not
just the reaction rate. This process is known as {\em 
composite operator renormalisation}. The
correct definition of the renormalised 2-point function
is therefore
\begin{equation}
\langle R_{\mu_1}R_{\mu_2}\rangle_{\rm R} = Z_2 \langle R_{\mu_1}R_{\mu_2}\rangle_{g\to g_R}, 
\end{equation}
where the amplitude $Z_2$ is chosen so that 
$\langle R_{\mu_1}R_{\mu_2}\rangle_{\rm R}$ is nonsingular in 
$\epsilon$:
\begin{equation}
Z_2 = 1+\frac{g_R}{2\pi\epsilon} + o(g_R^2).
\end{equation}
The prefactor, $Z_n$, of the $n^{\rm th}$ order correlation
function can be computed in a similar manner to the second 
order one. For example, in the loop expansion of the 
$3^{\rm rd}$ order correlation function, there are three 
diagrams containing singularities which are not removed
by coupling constant renormalisation. These are shown in
fig. \ref{fig-3pointFn}. For the $n$-point function there
are $\frac{1}{2}n(n-1)$ such diagrams. Each of these diagrams
contributes $\frac{g_R}{2\pi\epsilon}$ to the one-loop
expression for $Z_n$ so that : 
\begin{equation}
Z_n = 1 + \frac{1}{2}n(n-1) \frac{g_R}{2\pi\epsilon} + o(g_R^2).
\end{equation}

This situation is a bit more complicated than before. To extract the
scaling exponent we employ the technology of renormalisation
group (RG) which was not truly necessary to compute the scaling
of the density. Our discussion follows closely the presentation of 
\cite{binney}. The approach is based on the simple observation,
already made at the end of section \ref{sec-oneLoopAnswers},
that the $n^{th}$ order correlation function, $C^{(n)}(L_{\mu_1}\ldots L_{\mu_n})=
\langle R_{\mu_1}\ldots R_{\mu_n}\rangle$ does not depend on
the arbitrary length scale $L_\mu$, known in RG language as 
the {\em reference scale}. It immediately follows that
\begin{eqnarray}
\nonumber & &L_\mu\pd{}{L_\mu} \left( Z_n^{-1}(g_R)\, C^{(n)}_{\rm R}(L_{\mu_1}\ldots L_{\mu_n}, g_R, L_\mu) \right) = 0.
\end{eqnarray}
The $L_\mu$-dependence of the bracketed expression comes from three sources :
an explicit dependence of $C^{(n)}_{\rm R}$ on $L_\mu$, an implicit dependence
through $g_R(L_\mu)$ and  an implicit dependence through $Z_n(g_R(L_\mu))$.
We can thus write
\begin{eqnarray}
\label{eq-CS1} \left[ L_\mu\pd{}{L_\mu} + L_\mu\pd{g_R}{L_\mu}\pd{}{g_R} +
L_\mu\pd{Z_n}{L_\mu}\pd{}{Z_n}\right]Z_n^{-1}(g_R)\, C^{(n)}_{\rm
R}(L_{\mu_1}\ldots L_{\mu_n}, g_R, L_\mu) = 0,
\end{eqnarray}
where the partial derivative with respect to $L_\mu$ is now taken at fixed
$g_R$ and $Z_n$ whose dependences on $L_\mu$ are catered for by the additional
derivatives. This can then be arranged to give the equation :
\begin{eqnarray}
\label{eq-CS2} \left[ L_\mu\pd{}{L_\mu} + \beta(g_R)\pd{}{g_R} -
\gamma_n(g_R)\right] C^{(n)}_{\rm R}(L_{\mu_1}\ldots L_{\mu_n}, g_R, L_\mu) =
0,
\end{eqnarray}
where
\begin{eqnarray}
\nonumber \gamma(g_R) &=& L_\mu\pd{}{L_\mu} \left(\log Z_2(g_R)\right),\\
&=& \frac{g_R}{2\pi} + o(g_R^2),
\end{eqnarray}
and $\beta(g_R)$ is given by eq. (\ref{eq-beta}). By itself, this
equation just tells us how $C^{(n)}_{\rm R}$ varies with physically 
meaningless reference scale, $L_\mu$. However dimensional analysis provides 
extra information. Since the physical dimension of $C^{(n)}_{\rm R}$ is
$L^{-n d}$ it must satisfy an Euler equation \cite{binney}
\begin{equation}
\label{eq-Euler}
\left[\sum_{i=1}^n L_{\mu_i}\pd{}{L_{\mu_i}} + L_{\mu}\pd{}{L_\mu} +n
d\right]C^{(n)}_{\rm R}(L_{\mu_1}\ldots L_{\mu_n}, g_R, L_\mu) =0.
\end{equation}
Suppose we now rescale all lengths by some amount, $\Lambda$, by introducing
$\tilde{L}_{\mu_i}=\Lambda L_{\mu_i}$. Eq. 
(\ref{eq-Euler}) allows us to convert derivatives with respect to $L_\mu$
into derivatives with respect to $\Lambda$: 
\begin{equation}
L_\mu\pd{}{L_\mu} C^{(n)}_{\rm R}(\tilde{L}_{\mu_1}\ldots \tilde{L}_{\mu_n},
g_R, L_\mu)= -\left( \Lambda\pd{}{\Lambda} + n d\right)C^{(n)}_{\rm
R}(\tilde{L}_{\mu_1}\ldots \tilde{L}_{\mu_n}, g_R, L_\mu),
\end{equation}
so that eq. (\ref{eq-CS2}) can be written as
\begin{equation}
\label{eq-CS3} \left[ -\Lambda\pd{}{\Lambda} + \beta(g_R)\pd{}{g_R}  -n
d-\gamma_n(g_R)\right] C^{(n)}_{\rm R}(\tilde{L}_{\mu_1}\ldots
\tilde{L}_{\mu_n}, g_R, L_\mu) = 0.
\end{equation}
This equation is called the Callan-Symanzic (C-S) equation. It tells us something
physically useful, namely how the renormalised correlation function changes as 
we rescale its arguments by $\Lambda$. We wish to solve it in the limit
of large $\Lambda$. This can be done using the method of characteristics.
For $\Lambda=1$, $C^{(n)}_{\rm R}$ is given by the mean-field answer which
is valid for small values of the $\tilde{L}_{\mu_i}$'s, thus providing an
initial condition : 
\begin{equation}
C^{(n)}_{\rm R}(\Lambda=1, g_R=g_0) = g_0^{-\frac{n}{2}}(L_{\mu_1}\ldots
L_{\mu_n})^{-d}.
\end{equation}
For $d<2$, $\beta(g_R)=\epsilon g_R(1-\frac{g_R}{2 \pi \epsilon})$ and the 
characteristic equations are
\begin{eqnarray}
\dd{\Lambda}{s} &=& -\Lambda,\\
\dd{g_R}{s}  &=& \epsilon g_R(1-\frac{g_R}{2 \pi \epsilon}), \\
\dd{C^{(n)}_{\rm R}}{s} & =& \left(n d+\frac{1}{2}n(n-1)\frac{g}{2\pi}\right),
\end{eqnarray}
with the boundary conditions
\begin{eqnarray}
\nonumber \Lambda(g_0,s_0)& =& 1,\\
\label{eq:boundary}g_R(g_0,s_0)& = &g_0,\\
\nonumber C^{(n)}_{\rm R}(g_0,s_0) &= & g_0^{-\frac{n}{2}}(L_{\mu_1}
\ldots L_{\mu_n})^{-d}.
\end{eqnarray}

If we solve these equations, use the uniqueness of the characteristic 
curves to express $s_0$ and $g_0$ in terms of $\Lambda$ and $g_R$ and then
evaluate the solution at $s=0$, the answer can be found explicitly:
\begin{equation}
C^{(n)}_{\rm R}(\Lambda, g_R) = \left(\sqrt{\frac{1}{2\pi \epsilon}\left(1-\left(1-\frac{2 \pi \epsilon}{g_R}\right)\Lambda^{-\epsilon}\right)}\right)^n
\left(\Lambda L_{\mu_1}\ldots\Lambda L_{\mu_n}\right)^{-d}
\left(\frac{\Lambda^{-\epsilon}-\left(1-\frac{2 \pi \epsilon}{g_R}\right)\Lambda^{-\epsilon}}{1-\left(1-\frac{2 \pi \epsilon}{g_R}\right)\Lambda^{-\epsilon}}\right)^{\frac{1}{2}n(n-1)}.
\end{equation}
Taking $\Lambda\to\infty$ we conclude
\begin{equation}
C^{(n)}_{\rm R}(\tilde{L}_{\mu_1}\ldots \tilde{L}_{\mu_n}, g_R, L_\mu) \sim 
\prod_{i=1}^n L_{\mu_i}^{\frac{1}{2}\epsilon(n-1)}\sqrt{\frac{1}{2 \pi \epsilon}}\, \tilde{L}_i^{-d-\frac{1}{2}\epsilon(n-1)},
\end{equation}
independent of the value of $g_R$. This independence is the consequence of 
the presence of a fixed point of the $\beta$-function. All values of $g_R$
flow to the fixed-point value, $g^*=2\pi\epsilon$, leaving a universal answer
in the limit of large $\Lambda$. It remains to perform the inverse Laplace
Transform to find the scaling properties of the original mass-space correlation
functions.

To do this we note from eq. (\ref{eq-L}) that for large values of the $\tilde{L}_i$,
\begin{equation}
\tilde{L}_i = \left(\frac{J \tilde{\mu}_i}{D m_0}\right)^{-\frac{1}{d+2}}.
\end{equation}
It is then easy to perform the $n$ inverse Laplace Transforms with respect to
the $\tilde{\mu}_i$ to get
\begin{equation}
C^ {(n)}_{\rm R}(\tilde{m}_1\ldots \tilde{m}_n, g_R, L_\mu) \sim
\prod_{i=1}^n \tilde{m}_i^{-\frac{2 d +
2}{d+2}-\frac{\epsilon(n-1)}{2(d+2)}}.
\end{equation}

The mass scaling of $C^{(n)}_{\rm R}$ is therefore $m^{-\gamma_n}$ with
\begin{equation}
\gamma_n = n\frac{2d+2}{d+2} + \frac{n(n-1)\epsilon}{2(d+2)}+o(\epsilon^2).
\label{finalanswer}
\end{equation}
Note that $\gamma_n$ acquires a correction to the value
predicted from K41 theory signalling the breakdown of 
self-similarity in low dimensions.
This is the multiscaling curve against which 
we compared our numerical results in fig. \ref{fig-numerics}.

\subsection{Logarithmic Corrections in $d=2$}
\label{sec-RGdeq2}
In $d=2$ scale invariance is broken by the presence of logarithmic corrections
to the mean field scaling. For completeness, let us calculate the powers of
the logarithms acquired by the $C^{(n)}_R$'s. In $d=2$, $\beta(g_R)=-\frac{g_R^2}{2\pi}$ and the C-S equation, eq. (\ref{eq-CS3}), reads :
\begin{equation}
\label{eq-CS2D} \left[ -\Lambda\pd{}{\Lambda} - \frac{g_R^2}{2\pi}\pd{}{g_R}  -2 d-\frac{1}{2}n(n-1)\frac{g_R}{2\pi}\right] C^{(n)}_{\rm R}(\tilde{L}_{\mu_1}\ldots \tilde{L}_{\mu_n}, g_R, L_\mu) = 0.
\end{equation}
The initial condition is again given by the mean-field answer which in $d=2$
is
\begin{equation}
C^{(n)}_{\rm R}(\Lambda=1, g_R=g_0) = g_0^{-\frac{n}{2}}(L_{\mu_1}\ldots L_{\mu_n})^{-2}.
\end{equation}
The characteristic equations are
\begin{eqnarray}
\dd{\Lambda}{s} &=& -\Lambda, \\
\dd{g_R}{s} & =& -\frac{g_R^2}{2\pi},\\
\dd{C^{(n)}_{\rm R}}{s} & = &\left(2n+\frac{1}{2}n(n-1)\frac{g}{2\pi}\right),
\end{eqnarray}
with the boundary conditions as in eq. (\ref{eq:boundary}) with $d$
replaced by $2$.
These can again be solved explicitly at $s=0$  to give:
\begin{equation}
C_n^{\rm R}(\Lambda,g_R) = \prod_{i=1}^{n}
\sqrt{\frac{1}{g_R}+\frac{1}{2\pi}\log{\Lambda}}\, (\Lambda L_{\mu_i})^{-2} 
\left(\frac{2\pi}{g}\right)^{\frac{1}{2}(n-1)}
\left( \frac{2\pi}{g}+\log{\Lambda}\right)^{-\frac{1}{2}(n-1)},
\end{equation}
or, in terms of the rescaled lengths, $\tilde{L}_i$ : 
\begin{equation}
C_n^{\rm R}(\tilde{L}_1\ldots \tilde{L}_n,g_R,L_\mu) = \prod_{i=1}^{n}
\sqrt{\frac{1}{g_R}+\frac{1}{2\pi}\log{\frac{\tilde{L}_i}{L_{\mu_i}}}}\,
\tilde{L}_i^{-2}
\left(\frac{2\pi}{g}\right)^{\frac{1}{2}(n-1)}
\left(
\frac{2\pi}{g}+\log{\frac{\tilde{L}_i}{L_{\mu_i}}}\right)^{-\frac{1}{2}(n-1)}.
\end{equation}
The large mass limit corresponds to $\tilde{L}_i/L_{\mu_i} \to \infty$ in 
which case
\begin{equation}
C_n^{\rm R}(\tilde{L}_1\ldots \tilde{L}_n,g_R,L_\mu) \sim \prod_{i=1}^{n}
\sqrt{\frac{1}{2\pi}\log{\frac{\tilde{L}_i}{L_{\mu_i}}}}\, \tilde{L}_i^{-2} 
\left(\frac{2\pi}{g}\right)^{\frac{1}{2}(n-1)}
\left(\log{\frac{\tilde{L}_i}{L_{\mu_i}}}\right)^{-\frac{1}{2}(n-1)}.
\end{equation}
To recover the asymptotic behaviour in mass space, it is again necessary 
to take inverse Laplace transforms with respect to the $\tilde{\mu}_i$
as $\tilde{\mu}_i\to 0$. Recalling the definition, eq. (\ref{eq-L}) of $L_\mu$ 
we can write
\begin{equation}
C_n^{\rm R}(\tilde{m}_1\ldots \tilde{m}_n,g_R)\sim  \int \prod_{i=1}^{n} 
e^{-\tilde{m}_i\tilde{\mu}_i}d\tilde{\mu}_i\, \left(\sqrt{-\frac{1}{8\pi}\log{m_i\tilde{\mu}_i}}\, 
\sqrt{\frac{J\tilde{\mu}_i}{D m_0}}\right)^n \left(\frac{2\pi}{g}\right)^{\frac{1}{2}n(n-1)}
\left(-\frac{1}{4}\log{m_i\tilde{\mu}_i}\right)^{-\frac{1}{2}n(n-1)}.
\end{equation}
By introducing scaling variables $x_i=\tilde{m}_i \tilde{\mu}_i$ and keeping 
leading order terms in $\log{\tilde{m}_i/m_i}$ the asymptotic behaviour of this integral as
the $\tilde{m}_i\to\infty$ is shown to be
\begin{equation}
C_n^{\rm R}(\tilde{m}_1\ldots \tilde{m}_n,g_R) \sim C(g_R) \prod_{i=1}^n \left(\sqrt{\frac{J}{D}\,\log{\frac{\tilde{m}_i}{m_i}}}\,\tilde{m}_i^{-\frac{3}{2}}\right)\, \left(\log{\frac{\tilde{m}_i}{m_i}}\right)^{-\frac{1}{2}(n-1)}.
\end{equation}
The density, $n=1$, picks up a square root of a logarithm
coming from the renormalisation of the reaction rate. However the higher
order correlation functions pick up {\em additional} logarithmic corrections
which come from the anomalous dimension of the two point function. Note that
in $d=2$ the asymptotic behaviour retains some memory of the low mass
cut-offs, $m_i$. Furthermore the prefactor, denoted above by $C(g_R)$ remains
dependent on the value of $g_R$, unlike in $d<2$.

\section{Renormalised Smoluchowski Equation}
\label{sec-RSE}
Although the correspondence between constant kernel aggregation and the
$A+A\to A$ system is convenient for calculations and greatly simplifies the 
field theoretic description of the problem, the physics is sometimes obscured
by this mapping. In this section we show that the results for the average
mass density, derived in section \ref{sec-avgDensity} from a field
theoretic perspective, can also be interpreted in a more physically
transparent way using some heuristic arguments closely related to the
so-called Smoluchowski approximation used in \cite{krapivsky1994} to take into
account fluctuations in low dimensional heterogeneous annihilation.

The idea is as follows.  If we are interested in the average mass density, then
the only diagrams remaining in the field theory associated with the stochastic
rate equation, eq. (\ref{sre}), after all trees have been summed are those which 
renormalise the reaction rate. Therefore it should be possible to understand
the problem directly in terms of a standard Smoluchowski kinetic theory 
with a suitably modified kernel, without reference to the Laplace-transformed
field $R_\mu$. Furthermore, the stationary state of this renormalised
Smoluchowski theory can be found directly without need for any R.G. analysis
using a technique, known as the Zakharov Transformation, borrowed from the the
theory of wave turbulence.

The variation of the dimensionless reaction rate, $g$, as the length scale
$L_\mu$ is changed is determined by the $\beta$-functions computed in
section \ref{sec-betaFunction}. From eq. (\ref{eq-gdlt2}) and eq.
(\ref{eq-gdeq2}) we can obtain the renormalisation law for the physical
reaction rate, $\lambda_R(L_\mu) = g_R(L_\mu)\,L_\mu^{-\epsilon}$. Since the 
Laplace variable, $\mu$, can be thought of as an inverse mass, we can
make the heuristic substitution, 
\begin{equation}
L_\mu \sim \left(\frac{J}{D m_0}\right)^{-\frac{1}{4}}\,m^{\frac{1}{d+2}},
\end{equation}
to motivate the following mass-dependent reaction rates :
\begin{eqnarray}
\label{eq-renLambda} \lambda_R(m) &=& \frac{\lambda}{1+\frac{\lambda}{2 \pi
\epsilon} \left(\frac{J}{m_0
D}\right)^{-{\frac{\epsilon}{d+2}}}m^\frac{\epsilon}{d+2}}, \quad \mbox{$d<2$},\\
\label{eq-renLambda2}\lambda_R(m) &= &\frac{\lambda}{1+\frac{\lambda}{8 \pi} 
\ln \left(\frac{m}{m_0}\right)},\qquad \qquad \qquad \! \mbox{$d=2$}.
\end{eqnarray}
The renormalised Smoluchowski equation (RSE) is then obtained from the mean field
Smoluchowski equation by substituting the above mass-dependent reaction rate
for $\lambda$.  The density should therefore satisfy the following equation at 
large times
\begin{widetext}
\begin{eqnarray}
\nonumber  \frac{\partial N(m,t)}{\partial t} &=& 
\int_0^{\infty}dm_1 dm_2
\lambda_R(m) N(m_1,t) N(m_2,t) \delta(m-m_1-m_2)\\
\label{eq-RSE}&-&
\int_0^{\infty}dm_1dm_2
\lambda_R(m_2)N(m,t) N(m_1,t)\delta(m_2-m-m_1)\\
\nonumber&-&
\int_0^{\infty}dm_1dm_2
\lambda_R(m_1)N(m,t) N(m_2,t)\delta(m_1-m_2-m)\\
\nonumber
&+& \frac{J}{m_0}\,\delta(m-m_0).
\end{eqnarray}
\end{widetext}
The stationary state of this equation is best studied using the method of
Zakharov transformations as detailed in \cite{kontorovich2001,CRZ1}. For $d<2$ 
and $m \gg m_0$, $\lambda_R(m) = 2 \pi \epsilon (J/D)^\frac{\epsilon}{d+2} 
m^{-\frac{\epsilon}{d+2}}$. The constant flux solution obtained by 
applying the Zakharov transformation to eq. (\ref{eq-RSE}) with this kernel is
\begin{equation}
N(m) = c_K\, \left( \frac{J}{D}\right)^\frac{d}{d+2} m^{-\frac{2d+2}{d+2}}.
\end{equation}
The constant, $c_K$, can be calculated exactly. However since we have only
calculated the renormalised kernel heuristically, it does not make sense to do 
so at this point. Note that we recover the predictions of the K41 theory for the
large mass behaviour of the solution to the RSE.
Conversely, for small masses we see that $\lambda_R(m) = \lambda$. In this
limit, the Zakharov transformation gives the original mean field solution
for constant kernel aggregation,
\begin{equation}
N(m) =\sqrt{\frac{J}{4\pi\lambda}}m^{-\frac{3}{2}},
\end{equation}
as found already from our RG analysis. It is an easy task to show that both
the mean field and renormalised mean field density distributions are local
in the sense that the inertial range mass transfer in the stationary state 
does not depend strongly on the source (or the sink which removes large masses
if one is present). Therefore both are physically realisable solutions and
both carry the same flux, $J$, of mass from small masses to large. 
The complete stationary distribution for 
constant kernel aggregation should therefore exhibit a crossover from the
mean field solution to the renormalised mean field solution at a mass, $m_c$
which is given by $(J/D)^\frac{\epsilon}{d+2} 
m_c^{-\frac{\epsilon}{d+2}} \approx 1$.  We note that this crossover from a 
mean field spectrum to a fluctuation dominated one has been conjectured to 
occur in other turbulent systems, in particular in wave turbulence 
\cite{BNN2001}, although these systems do not readily lend themselves to such
systematic analysis.

In $d=2$, analysis of the RSE equation allows us to 
obtain the logarithmic correction to the mean field spectrum, again without
resort to R.G. arguments. Strictly speaking, the Zakharov transformation 
technology of \cite{kontorovich2001,CRZ1} only works for homogeneous kernels 
which leaves us with the question of what to do with the logarithm in eq.
(\ref{eq-renLambda2}). It turns out that the approach is easily adapted to 
extract the logarithmic correction to the spectrum at large masses. We give a 
brief outline here. In $d=2$, the renormalised interaction, eq.
(\ref{eq-renLambda2}), behaves for large masses as 
\begin{equation}
\lambda_R(m) \sim \frac{1}{\ln\left(\frac{m}{m_0}\right)}.
\end{equation}
After application of the Zakharov transformations described in 
\cite{CRZ1}, the stationary RSE, eq. (\ref{eq-RSE}), it becomes
\begin{eqnarray}
\nonumber  0 &=& 
\int_0^{\infty}dm_1 dm_2 \left[ \ln\left(\frac{m}{m_0}\right)^{-1} N(m_1)\, N(m_2)\right.\\
\nonumber&-& \ln\left(\frac{m^2}{m_2m_0}\right)^{-1} N(m)\, N(\frac{mm_1}{m_2})\, \left(\frac{m}{m_2}\right)^2\\
\nonumber&-& \left.\ln\left(\frac{m^2}{m_1m_0}\right)^{-1} N(m)\, N(\frac{mm_2}{m_1})\, \left(\frac{m}{m_1}\right)^2\right]\\
& &\delta(m-m_1-m_2). 
\end{eqnarray}
Let us now look for a solution of the form $N(m)=c_K\, \ln(m/m_0)^y m^{-x}$.
This substitution yields a rather messy expression which we analyse by 
introducing new integration variables, $\mu_1$, $\mu_2$ defined by 
$m_1=m\mu_1$, $m_2=m\mu_2$ and then expanding  the resulting expression as a 
power series in $\ln (m/m_0)^{-1}$. After some algebra one obtains :
\begin{eqnarray}
\nonumber  0 &=& 
c_K^2 m^{2-2x} \int_0^{\infty}d\mu_1 d\mu_2 (\mu_1\mu_2)^{-x}\left[ \ln\left(\frac{m}{m_0}\right)^{2y-1} \right. \\
\nonumber&-& \ln\left(\frac{m}{m_0}\right)^{2y-1} \mu_1^{2x-2} - \ln\left(\frac{m}{m_0}\right)^{2y-1} \mu_2^{2x-2}\\
&+&  \left.{\rm O}\left( \ln\left(\frac{m}{m_0}\right)^{-1}\right)\right]\ \delta(1-\mu_1-\mu_2). 
\end{eqnarray}
It is clear that the leading logarithms cancel out asymptotically as 
$m\to\infty$ if we choose $y=1/2$. The integrand then vanishes asymptotically
for $x=3/2$ as in the usual mean field case. The renormalised Kolmogorov
spectrum in $d=2$  for large masses is therefore
\begin{equation}
P(m) = c_K \sqrt{\ln \left(\frac{m}{m_0}\right)}\, m^{-\frac{3}{2}},
\end{equation}
as found from the RG analysis of section \ref{sec-RGdeq2}.

Let us close this section by discussing the connection between the 
renormalised Smoluchowski equation and the Smoluchowski approximation used
in \cite{krapivsky1994} to study the kinetics of heterogeneous annihilation. 
The essence of the argument used in \cite{krapivsky1994} is as follows. Consider
a heterogeneous system of annihilating particles where the reactants have a 
continuous distribution of diffusivities with the slower particles being less
probable than the faster ones. Now consider the reaction
between particles with diffusivity $D$ and ``slower'' particles with
diffusivity $D^\prime < D$. Since the slower particles are rare we can
neglect reactions between slow particles. We therefore estimate the effective 
reaction rate for the slow particles by considering each slow particle to be 
stationary in a uniform background cloud of faster particles and calculating 
the diffusive flux of fast particles reaching the slow particle from the
background cloud. One estimates the effective reaction rate in $d$ dimensions
at time $t$ to be $(D+D^\prime)^{d/2}t^{-1+d/2}$. Substitution of this 
effective reaction rate in the mean field rate equation constitutes the 
Smoluchowski approximation 
which greatly improves the estimate of the asymptotic decay rate of the
particle density in $d<2$. In the case at hand, the dynamics naturally generates a distribution of particle masses in which heavy particles are 
much rarer than the lighter ones. So, although in our model all particles
have equal diffusivity, $D$, in the reference frame of a heavy particle we can 
consider the main interaction to be with a uniform background cloud of
light particles. Thus one can envisage an effective reaction rate for
the large mass particles of $(2D)^{d/2}t^{-1+d/2}$. However in our
case the mass flux into the system is a constant, $J$, so that the system 
reaches a stationary state in which the natural unit of time for particles of 
mass $m$ is 
$t \sim (J^{-1}D^{-d/2} m)^{2/(d+2)}$. Thus the time dependent effective 
reaction rate of \cite{krapivsky1994} should be replaced in the stationary
state of the constant kernel aggregation problem with a mass dependent
reaction rate $\lambda_R(m) \sim D^{d/2} (J D^{d/2})^{-(d-2)/(d+2)} 
m^{(d-2)/(d+2)}$ which we recognise as the large mass behaviour of the
renormalised kernel, eq. (\ref{eq-renLambda}). Thus the large mass asymptotics of the
RSE corresponds to the correction of the mean field rate equation by
the Smoluchowski approximation. The field theory approach with which we
derived the original results allows unambiguous identification of the set of 
diagrams which have been summed to give this approximation.


\section{Conclusions}
\label{sec-conclude}
To summarise, the most important result of this work is the following fact : 
the steady state mass density PDF of cluster-cluster aggregation in the 
presence of a steady source of monomers exhibits non-trivial multiscaling in
dimensions less than or equal to 2. Technically, this means that the exponents,
$\zeta_n$, describing the large mass asymptotics of the density correlation 
functions, $C_n(m_1\ldots m_n) = \langle N_{m_1}\ldots N_{m_n} \rangle$, 
depend nonlinearly on the order, $n$, of the correlation function. Physically
it means that the system is in a regime where strong correlations (or in
this case, anticorrelations) between particles dominate the statistics putting
the problem outside of the domain of applicability of mean-field theory.
In the early sections of the article we established the presence of multiscaling
more-or-less rigorously by calculating the asymptotic scaling behaviour of the
$C_n$ for $n=0,1,2$ and remarking that they do not lie on one line. The
latter part of the article is devoted to detailing a nonrigorous derivation of 
$\zeta_n$ for general $n$ as an expansion in $\epsilon=2-d$ using 
renormalisation group techniques. To leading order in $\epsilon$, the scaling
curve is quadratic in $n$. In $d=2$ the power law corrections to the mean-field
exponents become logarithmic. Our results are interesting both from the 
perspective of particle systems and from the perspective of turbulence. In the
context of aggregation, the understanding of the detailed role of fluctuations
beyond their effect on the density is still embryonic. The effect of 
fluctuations on the density can be taken into account using the so-called
Smoluchowski approximation. In the closing section of the article we showed
how our approach is connected to this approximation. Whereas the Smoluchowski
approximation is only useful for the density, our methods work for arbitrary
correlation functions. The connection to the theory of turbulence 
stems from the analogy between the mass cascade in the stationary state of
cluster-cluster aggregation and the energy cascade in the stationary state
of homogeneous isotropic turbulence. Throughout the article we have tried to
emphasise both the usefulness and limitations of this analogy. Both systems
exhibit nontrivial multiscaling. In the case of turbulence, this multiscaling
has proven to be very difficult to understand quantitatively. One might hope 
that the analogy developed here might, if viewed in the correct way, provide
some insight. At present, however, this is merely an aspiration.


\begin{acknowledgments}
{CC acknowledges the support of Marie--Curie grant HPMF-CT-2002-02004. RR
acknowledges the support of NSF grant DMR-0207106.}
\end{acknowledgments}

\appendix

\section{Effective Hamiltonian of Mass Model.}
\label{app-MMHamiltonian}
A microstate of the mass model is given by the vector
of occupation numbers, $\vec{N}=\{ N_{\xv,m}\}_{\xv \in {\mathbf Z}^d, m \in
{\mathbf Z}^{+}}$. The probability measure ${\mathbf P}_{t}(\vec{N})$ 
on the space of microstates satisfies the Master Equation, which follows from
the dynamics of the mass model:
\begin{widetext}
 \begin{eqnarray*}
&&\pd{}{t}P_{t}(\vec{N})= \frac{D}{2d} \sum_{m}\sum_{\langle \xv, \xv' \rangle}\left( (N_{\xv,m}\!+\!1) P_{t}(\{\ldots n_{\xv,m}\!+1\!\ldots n_{\xv',m}\!-1\!\ldots\}) - N_{\xv,m} P_{t}(\vec{N})\right) \\
&&+ \frac{\lambda}{2}\sum_{\xv}\sum_{m=m_1+m_2}\left(
(N_{\xv,m_1}\!+\!1) (N_{\xv,m_2}+1) P_{t}(\{\ldots
N_{\xv,m_1}\!+1\!\ldots N_{\xv,m_2}\!+1\!\ldots
N_{\xv,m}\!-1\!\ldots\})\, - N_{\xv,m_1} N_{\xv,m_2} P_{t}(\vec{N})\right)\\
&&\nonumber \sum_{\xv}\sum_{m }\left(
\frac{J}{m_{0}}\delta(m-m_0)P_{t}(\{\ldots N_{\xv,m}-1 \ldots\})
 - \frac{J}{m_{0}}\delta(m-m_0)P_{t} (\vec{N}) \right).
\end{eqnarray*}
\end{widetext}
The first line of the Master Equation describes diffusion, second -
aggregation, the third - deposition. By "$\ldots$" we denote
components of the microstate, which are the same on the right hand
side as on the left hand side. Let $|P_{t} \rangle=\sum_{\vec{N}}
P_{t}(\vec{N})  \prod_{\xv,m}(\ad_{\xv,m})^{N_{\xv,m}} | 0
\rangle$ be the state vector corresponding to probability measure
$P_{t}$.

The evolution equation for $|P_{t} \rangle$ follows by differentiating
$|P_{t} \rangle$ with respect to time and using the master
equation to express the result back in terms of $|P_{t} \rangle$.
A simple calculation shows that
 \bea
 \partial_{t} |P_{t} \rangle =-H_{tm}|P_{t} \rangle,
 \eea
 where $H_{tm}$ is an effective Hamiltonian of the mass model
 given by : 
\begin{eqnarray}
H_{tm}=D\sum_{m}\sum_{\langle \xv, \xv' \rangle}
(\ad_{\xv,m}-\ad_{\xv',m})(a_{\xv,m}-a_{\xv',m})\nonumber\\
-\lambda \sum_{\xv} \sum_{m,m_{1}, m_{2}} \delta(m-m_{1}-m_{2}) \ad_{\xv,m} a_{\xv,m_{1}} a_{\xv,m_{2}} \nonumber \\
 + \lambda \sum_{\xv}
\sum_{m_{1},m_{2}} \ad_{\xv, m_{1}} \ad_{\xv, m_{2}}
a_{\xv,m_{1}}a_{\xv,m_{2}}\nonumber \\
-\sum_{\xv,m} \frac{J}{m_{0}} \delta(m-m_{0}) (\ad_{\xv,m}-1)
\label{eq-hamtm}
\end{eqnarray}
where $\langle \xv, \xv' \rangle$ denotes summation over pairs of nearest
neighbours.  Note that we replaced $\lambda$ with $2\lambda$ in order to 
simplify the notation. The Hamiltonian operator, eq. (\ref{eq-hamtm}), already 
includes the ``shift'' introduced to simplify the computation of correlation
functions as described in appendix \ref{sec-shiftExplanation}.

\section{Relation Between Correlation Functions of interacting particle systems and stochastic rate equations.}
\label{sec-shiftExplanation}
In this appendix, we establish the relation between interacting particle
systems and stochastic rate equations. 
To avoid unnecessary complications, we start with
an example of a zero-dimensional
system. We then show how the zero dimensional results get modified
in $d>0$ using the  example of the mass model.

The microstate of a zero dimensional particle system is determined
by the number of particles present at time $t$. Let the probability
that there are $N$ particles left at time $t$ be $P_{t}(N)$. If
the particle dynamics is Markovian, this probability satisfies the linear
master equation which is first order in time. The principle
step of Doi's map is to rewrite the master equation as an evolution
equation for a certain state vector in the Fock space spanned by
microstates of the particle system. The Fock space $\mathcal{F}$
can be introduced as follows. Let $| N \rangle$ be an element (a
"state") of $\mathcal{F}$, corresponding to a microstate with $N$
particles, $N=0,1,2,\ldots$. The state $| 0 \rangle$ is often
referred to as 'vacuum'. Then $\mathcal{F}$ is defined as a space
of all linear combinations of states $| N \rangle$ with complex
coefficients. Let $a^{\dagger}$ be the creation operator acting in
$\mathcal{F}$. By definition,
 \bea
 \ad | N \rangle =| N+1 \rangle
 \eea
 The annihilation operator, $a$, acting in $\mathcal{F}$ is defined as
follows:
 \bea
 [ a,\ad ] =1, \label{comr} \\
 a | 0 \rangle =0,
 \eea
 where $[ A , B ] \equiv AB-BA$ denotes the commutator of the 
operators $A,~B$. It then follows that
\be
a | N \rangle = N |N-1\rangle.
\ee
It is also convenient to define the left vacuum
$\langle 0 |$ (an element of the space dual to $\mathcal{F}$) by
means of the following relations:
 \bea
 \langle 0 | 0 \rangle =1, \nonumber\\
 \langle 0 | \ad =0.
 \eea
Consider the following element of $\mathcal{F}$:
 \bea
 | P_{t} \rangle =\sum_{N=0}^{\infty} P_{t}(N) | N \rangle.
 \eea
The state $| P_{t} \rangle$  can be used to generate
all probabilities $P_{t}(N)$, as $P_{t}(N)=\frac{1}{N!}\langle 0
| a^{N} | P_{t}\rangle$. The vector $| P_{t} \rangle$ satisfies an
equation, which follows from the master equation for
probabilities:
 \be
 \frac{d}{dt} |P_{t} \rangle = - H |P_{t} \rangle ,\label{schre}
 \ee
where $H$ is an evolution operator (Hamiltonian). The exact form
of the Hamiltonian follows from the microscopic rules of evolution.
For example, if particles undergo pairwise annihilations at
rate $\lambda$, then (see, for example, \cite{cardyhttp})
\be
H=-\lambda ( a^2-a^{\dagger~2} a^2).
\label{annih}
\ee
The first term on the right hand side of (\ref{annih}) describes
annihilation, while the second term comes from the "minus" term in the master
equation, which accounts for the probability of non-reaction. The formal
solution to (\ref{schre}) is
 \bea
 | P_{t} \rangle =e^{-tH} |P_{0} \rangle, \label{soln}
 \eea
where the state $|P_{0} \rangle$ is determined by the initial distribution
of particles. The problem of solving Master Equation is now
reduced to solving an effective Schroedinger Equation in imaginary
time, which can be often done using powerful methods of
quantum/statistical field theory.

Another important observation of Doi's theory is the averaging
formula. Let $Z_{t}(J)$ be the generating function of moments of
the probability distribution $P_{t}(N)$. By definition, $Z_{t}(J)
\equiv E(e^{JN_{t}}) =\sum_{0}^{\infty} e^{JN}P_{t}(N)$. Using
the definitions above, one can prove the following:
 \bea
 Z_{t}(J)= \langle 0 | e^{a} e^{J\ad a }| P_{t} \rangle.\label{genfun}
 \eea
To show this, expand the right hand side to give
\begin{eqnarray*}
 Z_{t}(J)&=&\sum_{N=0}^{\infty} \sum_{K=0}^{\infty} P_{t}(N)
\frac{J^{K}}{K !} \langle 0 | e^{a} (\ad a)^{K} | N \rangle \\
&=&\sum_{N=0}^{\infty} \sum_{K=0}^{\infty} \frac{J^{K}}{K !} N^{K}
P_{t} (N) \langle 0 | e^{a}| N \rangle\\
 &=&\sum_{N=0}^{\infty}
\sum_{K=0}^{\infty} \frac{J^{K}}{K !} N^{K} P_{t} (N) \\
&=&\sum_{N=0}^{\infty} e^{JN} P_{t} (N) \equiv E(e^{JN}). \\
\end{eqnarray*}

In the proof we used the fact that $\hat{N} \equiv \ad a$ is the
operator measuring the
number of particles, $\hat{N} | N \rangle = N | N \rangle$, and the identity
$e^a a^{\dagger ~N}=(\ad+1)^N e^a$.
The form of the correlation function on the right hand side of eq.~(\ref{genfun}) 
is not very suitable for practical computations due to the presence of
the shift operator $e^{a}$ inside the brackets. This problem can be overcome
by commuting $e^{a}$ to the right and using $e^a | 0 \rangle =| 0 \rangle$.
It follows from the commutation relations [eq.~(\ref{comr})]
that $e^{a} O(\ad, a)=O(\ad+1, a)$ for any operator $O (\ad, a)$.
Using this fact one finds that
 \bea
 Z_{t} (J) = \langle 0 | e^{J(\ad+1)a} e^{-t\tilde{H}}| \tilde{P}_{0}\rangle,
 \eea
where $\tilde{H} (\ad, a) = H (\ad +1, a )$, $|\tilde{P}_{0}
\rangle = e^a | P_{0} \rangle$. The expression $\langle 0 |
e^{J(\ad+a)a}$ can be simplified further. Note that $[\ad a ,a]=-a$. In
other words, the operators $a$ and $\ad a$ form a basis of a Lie
algebra isomorphic to a subalgebra of $sl(2)$ consisting of upper
triangular matrices. Consequently, the Campbell-Hausdorff formula
implies that
 \bea
e^{J(\ad+a)a} = e^{J\ad a} e^{f(J) a}, \label{ch}
 \eea
where $f(J)$ is a function to be determined. Differentiating both
sides of eq. (\ref{ch}) with respect to $J$, commuting all operators
multiplying exponents in the derivatives to the right, and
comparing both sides of the resulting equality, we find a
differential equation for $f$:
 \bea
 f'(J)=f(J)+1,
 \eea
which should be solved with the boundary condition $f(0)=0$. The
answer is $f(J)=e^J -1$. Taking into account that 
$\langle 0 | e^{J\ad a}=\langle 0 |$, we find
\bea
 Z_{t} (J) = \langle 0 | e^{(e^{J}-1)a} e^{-t\tilde{H}}| \tilde{P}_{0}\rangle,
 \eea
The right hand side of this relation can be rewritten as a path
integral using the Trotter formula. Assume for simplicity that initial
probability distribution of the number of particles $P_{0} (t)$ is
Poisson with intensity $N_{0}$. Then,
 \bea
 Z_{t} (J)=\int \int \prod_{\tau} da(\tau )da(\tau)
 e^{(e^{J}-1)a(t)}e^{-S_{eff} (t)},
 \eea
where the integration is performed over the space of complex paths
$(\bar{a}(\tau), a (\tau))_{\tau \in [0,t]}$ and
 \bea
 S_{eff}(t)=\int_{0}^{t} d\tau \bigg( \bar{a}(\tau)
 \partial_{\tau} a(\tau)+\tilde{H}(\bar{a}(\tau),
 a(\tau))-N_{0}\bar{a}(\tau)\bigg) \label{action0d}
 \eea
 is an effective 'action functional'. Here
$\tilde{H}(\bar{a}(\tau),a(\tau))$ is the symbol of operator $\tilde{H}
(\ad, a)$, which is assumed to be normally ordered.
We conclude that
 \be
 E(e^{J N_{t}})_{P_{t}}=\langle e^{(e^J -1)a(t)}\rangle_{S_{eff}(t)},
 \label{thm}
 \ee
where we stressed by our notations, that the averaging in the 
left hand side of eq.~(\ref{thm}) is
performed over the space of microstates using probability distribution $P_{t}$,
while averaging in the right hand side is performed over the space of
all paths $(\bar{a}(\tau), a (\tau))_{\tau \in [0,t]}$ using the functional 
measure $e^{-S_{eff}(t)}$.
Differentiating both sides of eq.~(\ref{thm}) with respect to
$\theta=e^J-1$ and setting $\theta=0$ we find that moments of
$a(t)$ correspond to $factorial$ moments of $P_{t}$:
 \be
E\big( N_{t}(N_{t}-1)(N_{t}-2)...(N_{t}-k+1) \big)_{P_{t}}\nonumber \\ =\langle
a(t)^{k}\rangle_{S_{eff}(t)},k=1,2,3,\ldots
 \label{cor}
 \ee
In particular $E(N_{t})=\langle a(t) \rangle, E(N_{t}^2)=\langle
a(t)^2+a(t) \rangle$, etc.

Eq. (\ref{cor}) allows one to capture strong non-mean field behaviour
of an interacting particle system even if the field theory characterised by
eq.~(\ref{action0d}) is well approximated by mean field theory. The
simplest example is as follows. Assume that 
$\langle a(t)^{k} \rangle \approx \langle
a(t) \rangle^{k}$, but $E(N_{t})=\langle a(t) \rangle \ll1$. Using these
assumptions and eq.~(\ref{cor}) to evaluate moments of $N_{t}$, we find
that $E(N^{k})=E(N)$, which is essentially non-mean field
behaviour.

For particle systems with pairwise local interactions, 
eqs.~(\ref{thm}) and (\ref{cor}) can be formulated in terms of
stochastic differential equations thus avoiding references to
non-rigorous path integral constructions. To illustrate this,
consider the reaction $A+A\rightarrow \emptyset$ in zero dimensions.
The effective Hamiltonian is given by eq.~(\ref{annih}),
$\tilde{H}\equiv H(\ad+1,a)=\lambda ( 2\ad a^2+a^{\dagger~2}
a^2)$. The corresponding functional integral measure can be rewritten
using the Hubbard-Stratonovich identity as 
 \bea
 \lefteqn{ e^{-\int_{0}^{t} d\tau \bigg( \bar{a}(\tau)
\partial_{\tau} a(\tau)+\lambda ( 2\ad a^2+a^{\dagger~2}
a^2) -N_{0}\bar{a}(\tau)  \bigg)} =\prod_{\tau'} d\xi (\tau')
e^{-\frac{1}{2}\int_{0}^{t}d\tau \xi^{2}(\tau)}\nonumber} \quad\quad\\
&& \times  ~e^{-\int_{0}^{t}d\tau \bigg( \bar{a}(\tau)
\partial_{\tau} a(\tau)+2\lambda \ad a^2 -N_{0}\bar{a}(\tau)
-i\sqrt{2\lambda}\xi(\tau)\ad(\tau )a(\tau)  \bigg) }.
 \label{hastr}
 \eea
The field $\xi (t)$ is standard Gaussian white noise.
Note that the expression in the exponent in the second line of
eq.~(\ref{hastr}) is linear in the field $\ad$, which allows it to
be integrated out. This results in a $\delta$-functional with an
argument
 \be
\partial_{\tau} a(\tau)+2\lambda a^2 -N_{0}\delta(\tau)
-i\sqrt{\lambda}\xi(\tau)a(\tau)=0, \label{ra0d}
 \ee
in which we recognise a rate equation of $A+A \rightarrow
\emptyset$ augmented by imaginary multiplicative noise term
(Lee-Cardy equation).

We can now interpret eqs.~(\ref{thm}) and (\ref{cor}) as
follows. Let $N_{t}$ be the number of particles left after time
$t$ in the system of annihilating particles. Let $a(t)$ be the
solution to stochastic differential eq.~(\ref{ra0d}). Then
eqs.~(\ref{thm}) and (\ref{cor}) are true, given that $\langle
...\rangle_{S_{eff}}$ denotes averaging over the white noise
$\xi(t)$ \footnote{In deriving this statement one also has to
check that the value of the determinant of the functional
derivative of the left hand side of eq.~(\ref{ra0d}) with respect to $\phi$
is one. This is true in forward time regularization, which we
always implicitly assume.}. In the stated form, the correspondence
between the Markov chain describing $A+A\rightarrow \emptyset$ system
and the Lee-Cardy stochastic partial differential equation can be
proven rigorously and extended to a large class of interacting
particle systems in $d\geq 0$ \cite{TZ}.

Here we will only give a path integral derivation of the relation
between correlation functions in the mass model in $d>0$ and
correlation functions of Stochastic Smoluchowski equation, eq. (\ref{sse}).

First, consider the case of discrete space and mass. A 
microstate of the mass model is specified by stating the number of
particles of a given mass at a given site. In other words, the
microstate is a vector $\{ N_{\xv,m} \}_{\xv \in Z^d, m \in
Z_{+}}$ with non-negative integer components. The operator of mass
distribution is
 \bea
 \widehat{N}_{\xv,m}=\ad_{\xv,m} a_{\xv, m},
 \label{mdo}
 \eea
where $\ad_{\xv,m}$  and $a_{\xv, m}$ are creation and
annihilation operators satisfying the following commutation
relations:
 \bea
[ \ad_{\xv, m}, a_{\xv', m'} ]=\delta_{\xv, \xv'} \delta_{m,m'}.
 \eea
Similarly to the zero dimensional case considered above, the
generating functional for mass distribution correlation functions
can be written as follows:
 \bea
 Z_{t}[\vec{J}]&\equiv& E\bigg( e^{\sum_{\xv, m} J_{\xv, m}
N_{t,\xv,m}} \bigg), \nonumber \\ 
&=& \langle 0 | e^{\sum_{\xv, m}
a_{\xv, m}} e^{\sum_{\xv, m} J_{\xv, m}
 \widehat{N}_{\xv,m}} e^{-H_{\rm MM}} |P_{0}\rangle, \label{gftm}
 \eea
where
 \bea
H_{\rm MM}&=&D\sum_{m}\sum_{\langle \xv, \xv' \rangle}
(\ad_{\xv,m}-\ad_{\xv',m})(a_{\xv,m}-a_{\xv',m})\nonumber\\
&&-\lambda \sum_{xv} \sum_{m,m_{1}, m_{2}} \delta(m-m_{1}-m_{2}) \ad_{xv,m}
a_{\xv,m_{1}} a_{\xv,m_{2}} \nonumber \\
&& + \lambda \sum_{\xv}
\sum_{m_{1},m_{2}} \ad_{\xv, m_{1}} \ad_{\xv, m_{2}}
a_{\xv,m_{1}}a_{\xv,m_{2}}\nonumber \\
&& -\sum_{\xv,m} \frac{J}{m_{0}} \delta(m-m_{0}) (\ad_{xv,m}-1)
 \label{hamtm}
 \eea
  is an effective Hamiltonian of the mass model and $P_{0}$
is an initial probability measure on the space of microstates. An outline of the
derivation of  $H_{\rm MM}$ is presented in appendix \ref{app-MMHamiltonian}.

As before, we need to commute the projection operator $exp{\sum_{\xv,
m} a_{\xv, m}}$ to the right. Computations, which are completely
analogous to those performed in $0$-dimensional case give:
 \bea
Z_{t}[\vec{J}] = \langle 0 | e^{\sum_{\xv, m} (exp( J_{\xv,
m})-1) a_{\xv,m}}  e^{-\tilde{H}_{\rm MM}} |\tilde{P}_{0}\rangle,
\label{ztm}
 \eea
where $\tilde{H}_{\rm MM}$ is obtained from $H_{\rm MM}$ by the shift
$\ad_{\xv,m} \rightarrow \ad_{\xv, m}+1$ and
$|\tilde{P}_{0}\rangle =e^{\sum_{\xv, m} a_{\xv, m}}
|P_{0}\rangle$. The right hand side of eq.~(\ref{ztm}) can be re-written
in path integral form in the conventional way. A lengthy but
straightforward calculation shows that
 \bea
 Z_{t}[\vec{J}]&=&\int \prod_{\xv',m', \tau'} d\bar{\phi}(\xv',m',\tau')
d\phi (\xv',m',\tau')\nonumber \\
&& e^{\sum_{\xv, m} (exp( J_{\xv, m})-1) \phi (\xv,m,t)} e^{-S_{\rm MM}(t)},
\label{fi}
 \eea
 where
 \bea
 S_{\rm MM}(t)&\equiv&
\int_{0}^{t}dt\left(\sum_{\xv,m} \bar{\phi}(\xv,m,t)
\partial_{t}\phi(\xv,m,t)+\tilde{H}_{\rm MM}[\bar{\phi},\phi]  \right)\nonumber\\
&=&\int_{0}^{t}dt \sum_{\xv,m} \bigg( \bar{\phi}(\xv,m,t)\bigg[
\left( \frac{\partial}{\partial t} -D \Delta \right) \phi(\xv,m,t)
\nonumber \\
&&- \la \sum_{m'}  \phi(\xv, m',t) \phi(\xv, m-m',t) \nonumber \\
&& + 2\la \phi(\xv, m,t)\sum_{m'}\phi(\xv,m',t)-\frac{J}{m_{0}}\delta_{m,m_{0}}
\bigg] \nonumber\\
&& + \lambda \sum_{m'}
\bar{\phi}(\xv,m,t)\phi(\xv,m,t)\bar{\phi}(\xv,m',t)\phi(\xv,m',t)
\bigg), \label{sefftm}
 \eea
 where $\Delta$ is a discrete Laplacian and $J$ is the rate of input of mass
into the system.
Note that the expression in square brackets in the right hand side of
eq.~(\ref{sefftm}) is just the constant kernel Smoluchowski equation.
The last term in $S_{\rm MM}$ accounts for all correlation effects.
The exponential of this term can be rewritten using the
Hubbard-Stratonovich transformation as follows:
 \bea
 e^{-\lambda \int_{0}^{t} dt \sum_{\xv} \left( \sum_{m} \bar{\phi}(\xv,m,t)
\phi(\xv,m,t)
\right)^2}& =&\int \prod_{\xv',\tau'} d\xi(\xv',\tau' )
e^{-\frac{1}{2}\int_{0}^{t} dt
\sum_{\xv}
\xi(\xv,t)^2} \nonumber \\ 
&&e^{i\sqrt{2\lambda} \int_{0}^{t} dt
\sum_{\xv,m}
\xi(\xv,t)\bar{\phi}(\xv,m,t)\phi(\xv,m,t)}. \label{hstm}
 \eea
Note that the field $\xi$ is Gaussian, uncorrelated both in space
and time. Using eq.~(\ref{hstm}), the functional measure of
integration in eq.~(\ref{fi}) can be rewritten in the form
 \be
e^{-S_{\rm MM}}=\int \prod_{\xv', \tau'} d\xi(\xv', \tau')
e^{-\frac{1}{2}\int_{0}^{t}dt
\xi^2(\xv,t)} e^{-\int_{0}^{t} \sum_{\xv,m}
\bar{\phi}(\xv,m,t)L[\phi, \xi] },
\label{eq:29}
 \ee
 where
 \bea
 L[\phi, \xi]&=&\left(\frac{\partial}{\partial t} -D \Delta \right)
\phi(m) - \la \sum_{m'=0}^{m}  \phi(\xv,m',t) \phi(\xv,m-m',t)
\nonumber\\
&& + 2\la \phi(\xv,m,t) \sum_{m'=0}^{\infty}\phi(\xv,m',t) 
-\frac{J}{m_{0}}\delta_{m,m_{0}} -i\sqrt{2\lambda}\phi(m) \eta(\xv,t). 
\label{elagr}
 \eea
The  exponent in the right hand side of eq.~(\ref{eq:29}) is linear
in $\bar{\phi}$. Hence the path integral over fields $\bar{\phi}$,
$\phi$ and $\xi$ localises to paths satisfying Euler-Lagrange
equation
 \be
 \frac{\delta}{\delta \bar{\phi}(\xv,m,t)} \int_{0}^{t}dt \sum_{\xv,m}
\bar{\phi}(\xv,m,t) L[\phi, \xi],
 \ee
or
\be
 L[\phi, \xi]=0,\label{dsse}
\ee
which is a discrete version of the Smoluchowski equation, eq. (\ref{sse}).
For the sake of clarity, we restate the result concerning the
relation between the mass model and stochastic Smoluchowski
equation here. In order to calculate the generating functional of
density correlation functions in the mass model [eq.~(\ref{gftm})], 
one has to solve the 
Stochastic Smoluchowski equation, eq.(\ref{dsse}), for $\phi[\xi](\xv,m,t)$, then
average
 $e^{\sum_{\xv, m} (exp( J_{\xv,
m})-1) \phi[\xi](\xv,m)}$ with respect to Gaussian white noise $\xi$.
In other words,
 \be
 Z_{t}(\vec{J})=E\bigg( e^{\sum_{\xv, m} (exp( J_{\xv,
m})-1) \phi[\xi](\xv,m)} \bigg)_{\xi} \label{dmrel}
 \ee

Our final task is to discuss the modification and consequences of
eq.~(\ref{dmrel}) in the continuous limit. The latter is taken
according to the following set of rules:
\bea
\xv &\rightarrow& \frac{\xv}{a_{x}},\nonumber\\ 
m &\rightarrow & \frac{m}{a_{m}},\nonumber\\
\phi(\frac{\xv}{a_{x}},\frac{m}{a_{m}},t)&\rightarrow& \frac{1}{a_{x}^d
a_{m}} \phi (\xv, m, t),\nonumber\\
N(\frac{\xv}{a_{x}},\frac{m}{a_{m}},t)&\rightarrow &\frac{1}{a_{x}^d a_{m}} 
N (\xv, m, t),\nonumber\\
D &\rightarrow& \frac{D}{a_{x}^2},\nonumber\\
J& \rightarrow& \frac{Ja_{x}^d}{ a_{m}},\nonumber\\
m_{0}& \rightarrow& \frac{m_{0}}{a_{m}},\label{cl}
\eea
where $a_{x}$ and $a_{m}$ are lattice cut-offs in $\xv$- and
$m$-spaces respectively. The continuous limit is obtained by
performing replacements eq.~(\ref{cl}) in eqs.~(\ref{elagr}) and
(\ref{dmrel}) and taking lattice cut-offs to zero while keeping other 
parameters fixed.  As a result one recovers the stochastic Smoluchowski 
equation (SSE) [eq.~(\ref{sse})].

Note that the continuous field theory equivalent to SSE is
renormalizable in dimensions two and less, therefore eq.~(\ref{cl})
is justified in these dimensions only.
The continuous counterpart of eq.~(\ref{dmrel}) is
  \be
 Z_{t}(\vec{J})=E\bigg( e^{\int\int d\xv dm (exp( J(\xv, m))-1)
 \phi[\xi](\xv,m)} \bigg)_{\xi} \label{mrel}
 \ee

\section{Expression of probability of multi-particle configurations in terms of solutions to SSE.}
\label{app-SSE}
In this appendix, we show how probability of multiparticle configurations can
be calculated from SSE.
Equation~(\ref{mrel}) leads to the following relation between correlation
functions:
\bea
E\bigg( \phi(\xv,m,t) \bigg)_{\xi} &=&
E\bigg( N_{t}(\xv,m) \bigg), \nonumber\\
E\bigg( \phi(\xv_{1},m_{1},t) \phi(\xv_{2},m_{2},t) \bigg)_{\xi}
&=&E\bigg( N_{t}(\xv_{1},m_{1})N_{t}(\xv_{2},m_{2})\nonumber \\
&&-\delta^{d}(\xv_{1}-\xv_{2})\delta(m_{1}-m_{2})N_{t}(\xv_{1}, m_{1}) 
\bigg)_{\xi},
\label{moms}
\eea
and so on. Suppose that we are interested in the statistics of the total
number of particles $\Delta N_{t}(\xv,m)$ in a volume element
$\Delta V$ centred around $\xv$ with masses in the interval
$[m,m+\Delta m]$. In terms of the local mass distribution,
\be
\Delta N_{t}(\xv,m)=\int_{\Delta V} d^{d}x' \int_{m}^{m+\delta m} dm'
N_{t}(\xv',m').
\ee
Let
\be
\Delta \phi_{t}(\xv,m)=\int_{\Delta V} d^{d}x' \int_{m}^{m+\delta m}
dm'
\phi_{t}(\xv',m').
\ee
Integrating eq.~(\ref{moms}) with respect to mass and space, we find 
 \bea
 E\bigg( \Delta \phi (\xv, m,t)\bigg)&=&E\bigg( \Delta N_{t}(\xv, m,
t)\bigg) \nonumber\\
 E\bigg( \Delta \phi^2 (\xv, m,t)\bigg)&=&E\bigg( \Delta N_{t}(\xv, m,
t)(\Delta N(\xv,t,m)-1)\bigg)  \nonumber\\
 &\cdots& \nonumber \\
 E\bigg( \Delta \phi^n (\xv, m,t)\bigg)&=&E\bigg( \prod_{k=0}^{n-1}(\Delta
 N(\xv,t,m)-k)\bigg),\label{fmd}
 \eea
which is a multi-dimensional counterpart of $0$-dimensional result
of eq. (\ref{cor}).

In this paper we study scaling properties of probability of
finding multiple particles of large mass in $\Delta V \Delta m$.
Density of such particles is low. Thus, factorial moments entering
the right hand side of eq.~(\ref{fmd}) can be estimated in the limit of large
mass $m$ as follows:
\begin{eqnarray}
E\bigg( \prod_{k=0}^{n-1}(\Delta
 N(\xv,t,m)-k)\bigg) &\equiv &
\sum_{p=n}^{\infty}\bigg( \prod_{k=0}^{n-1}(p-k) \bigg) 
Prob\bigg(\xv_{1},
\ldots, \xv_{p} \in \Delta V; m_{1}, \ldots,
m_{p} \in [m,m+\Delta m] \bigg) \nonumber \\ 
&\approx & n! Prob\bigg(\xv_{1},
  \ldots, \xv_{n} \in \Delta V; m_{1}, \ldots,
 m_{p} \in [m,m+\Delta m] \bigg)
\end{eqnarray}

Combining this result with eq. (\ref{fmd}), we obtain desired relation
between probabilities of multi-particle configurations and moments
of solutions to SSE in the limit of large masses:
 \be
 Prob\bigg(\xv_{1},
  \ldots, \xv_{n} \in \Delta V; m_{1}, \ldots,
 m_{p} \in [m,m+\Delta m] \bigg)=
 \frac{1}{n!}E\bigg( \Delta \phi^n (\xv, m,t)\bigg)_{\xi}
 \ee

\bibliography{ref}
\end{document}